\documentclass[usegraphicx,usenatbib]{mn2e}
\usepackage{amsmath}
\usepackage{amssymb}
\usepackage{mathptmx}
\usepackage{subfigure}
\usepackage{epsfig}
\usepackage{bm}
\usepackage[usenames,dvipsnames]{color}
\definecolor{red}{rgb}{1,0,0}

\renewcommand{\descriptionlabel}[1]%
  {\hspace{\labelsep}\textbf{#1}}

\title[DIA: A Spatially Varying Photometric Scale Factor and Other Considerations]
      {Difference Image Analysis: Extension to a Spatially Varying Photometric Scale Factor and Other Considerations}

\author[D.M. Bramich et al.]
  {D. M. Bramich$^{1}$\thanks{E-mail: dbramich@eso.org, dan.bramich@hotmail.co.uk},
   Keith Horne$^{2}$,
   M. D. Albrow$^{3}$,
   Y. Tsapras$^{4,5}$,
   C. Snodgrass$^{6}$,
   \newauthor
   R. A. Street$^{4}$,
   M. Hundertmark$^{2}$,
   No\'{e} Kains$^{1}$,
   A. Arellano Ferro$^{7}$,
   R. Figuera Jaimes$^{7}$
   \newauthor
   and Sunetra Giridhar$^{8}$
  \medskip
  \\$^{1}$European Southern Observatory, Karl-Schwarzschild-Stra$\beta$e 2, 85748 Garching bei M\"{u}nchen, Germany
  \\$^{2}$SUPA Physics \& Astronomy, North Haugh, St Andrews, KY16 9SS, Scotland, UK
  \\$^{3}$Department of Physics and Astronomy, Private Bag 4800, University of Canterbury, New Zealand
  \\$^{4}$Las Cumbres Observatory, 6740B Cortona Dr, Suite 102, Goleta, CA 93117, USA
  \\$^{5}$School of Physics and Astronomy, Queen Mary University of London, Mile End Road, London E1 4NS, England
  \\$^{6}$Max Planck Institute for Solar System Research, Max-Planck-Str. 2, 37191 Katlenburg-Lindau, Germany
  \\$^{7}$Instituto de Astronom\'ia, Universidad Nacional Aut\'onoma de M\'exico, M\'exico
  \\$^{8}$Indian Institute of Astrophysics, Koramangala 560034, Bangalore, India
  }

\begin{document}

\date{Accepted 2010 August ???. Received 2010 August ???; Submitted 2010 August ???}

\pagerange{\pageref{firstpage}--\pageref{lastpage}} \pubyear{2010}

\maketitle

\label{firstpage}

\begin{abstract} 
We present a general framework for matching the point-spread function (PSF), photometric
scaling, and sky background between two images, a subject which is commonly referred to as difference
image analysis (DIA). We introduce the new concept of a spatially varying photometric scale factor which will be important
for DIA applied to wide-field imaging data in order to adapt to transparency and airmass variations across the field-of-view.
Furthermore, we demonstrate how to separately control the degree of spatial
variation of each kernel basis function, the photometric scale factor, and the differential sky background.
We discuss the common choices for kernel basis functions within our framework, and we introduce
the mixed-resolution delta basis functions to address the problem of the size of the least-squares problem
to be solved when using delta basis functions. We validate and demonstrate our algorithm on simulated and real data.
We also describe a number of useful optimisations
that may be capitalised on during the construction of the least-squares matrix and which have not been reported
previously. We pay special attention to presenting a clear notation for the DIA equations which are set out in a way
that will hopefully encourage developers to tackle the implementation of DIA software.
\end{abstract}

\begin{keywords}
methods: statistical -
techniques: image processing -
techniques: photometric
\end{keywords}

\section{Introduction}
\label{sec:introduction}

Difference image analysis (DIA) aims to measure changes, from one image to another, in the objects that make up a scene. In astronomy, the 
objects are typically point sources changing in brightness or moving on the sky. Astronomical images are formed on a discrete detector array, 
after the sky scene suffers attenuation, geometrical distortion and blurring by the atmosphere and optics, superimposed on a sky 
background, and corrupted by detector noise. All of these effects are to different degrees non-uniform across the scene and variable on a 
variety of timescales. Furthermore, pairs of images of the same scene may suffer small misalignments in position or scale, or gross rotational
misalignments.

The changes in object properties that we wish to measure are thus entangled with changes in the sky-to-detector, or scene-to-image, 
transformation. A residual difference image, formed by simple subtraction of one image from another, is generally dominated by changes in the
transformation. To extract the astronomical information, we must accurately model the changes in astrometry, throughput, background, and blurring
between the two images. We may then make corrections to match these effects from one image to another and subtract to form ``cleaner'' difference
images, or we may model the original images including changes in both object properties and image transformations. While
current DIA techniques are based on the former approach, we advocate the latter.

The model adopted to represent changes in the scene-to-image transformation must include the following differential (or corrective) components:
\begin{itemize}
\item{A coordinate transformation between the coordinate systems of each image to correct for
      image misalignments and/or differences in distortion.}
\item{A photometric scaling that corrects for the changes in the attenuating effects of the atmosphere
      (and possibly the telescope optics) and differences in exposure time.}
\item{A background offset that corrects for changes in the sky background emission.}
\item{A convolution transformation that corrects for the changes in the image point-spread function
      (PSF) as a result of changes in atmospheric conditions and/or the telescope optics (e.g. focus
      changes).}
\end{itemize}
Note that all of these components model differential corrections, not absolute values
(e.g. the convolution transformation models the {\it change} in PSF shape between images,
not the PSF itself).

The state of the art in DIA includes the components for photometric scaling, sky background offsets,
and PSF convolution in the DIA modelling process. Recent developments (\citealt{bra2008}, from now on B08) also
include fractional pixel translations in the model. Other image misalignments (rotation, scale, shear, distortion)
are corrected by pre-registering the images before application of DIA, usually involving image resampling.

The framework for the current approach to DIA was introduced by \citet{ala1998} (from now on A98) for matching
a reference image to a target image. The convolution kernel (including the photometric scaling) 
to be applied to the reference image is decomposed into a set of basis functions, and the differential background
offset is included as a polynomial of the image coordinates, which converts the problem of finding the
corrective components to a standard linear least-squares formulation.
A follow-up paper by \citet{ala2000} (from now on A00) showed how the spatial variation of the convolution
kernel can be modelled by multiplying the kernel basis functions by polynomials of the image coordinates.
The kernel basis functions chosen by A98 and A00 are Gaussians of different widths, modified by polynomials
of the kernel coordinates. The user must specify the number of Gaussian basis functions to be employed, their
associated widths, and the degrees of the modifying polynomials. However, the optimal choice of parameters
for generating the kernel basis functions is not obvious, although some investigation into this matter
has been performed (\citealt{isr2007}).

It is clearly desirable to find a set of kernel basis functions that are inherently simple, thereby
being specified by a minimal parameter set, and yet that can model the kernel with sufficient flexibility.
A step towards this paradigm was made by B08 with the proposed representation of the kernel
as a discrete pixel array where the kernel pixel values are solved for directly. This approach limits
the requirements on the user to specifying the kernel size (and shape), and the kernel model is maximally flexible
in modelling the most complicated convolution kernels (e.g. telescope jumps).
B08 show that the new formulation is capable of modelling fractional pixel translations as part
of the convolution kernel, thereby relaxing the requirement on image registration such that images need
only be aligned to the nearest pixel before application of DIA.
Spatial variation of the kernel is handled by interpolation of kernel and differential
background solutions on a grid.

Soon after B08, \citet{mil2008} (from now on M08) specified a set of kernel basis functions built from
delta-functions centred at different kernel coordinates. This choice of basis functions leads to a solution that happens to be equivalent to
the B08 solution (see Section~\ref{sec:delta_basis}), but it is specified such that it fits into the A98 framework of equations.
M08 also included a polynomial spatial variation of the delta-function coefficients to model the kernel spatial variation.
\citet{qui2010} ``rediscovered'' the M08 work, but failed to impose any control on the
photometric scaling while also fixing the value of the central kernel pixel, leading to a sub-optimal kernel model
that cannot freely model fractional pixel translations.

The choice of kernel basis functions in the A98 framework is fully down to the developer/user. While the delta-function
basis (or delta basis for short) is very compelling, the number of free parameters grows quickly with the adopted kernel size.
Hence it makes sense to choose some coarser functions in the outer part of the kernel where there is little
variation or signal/amplitude. \citet{alb2009} introduce the idea of binned kernel pixels in the outer part of
the kernel, which greatly reduces the number of kernel parameters, and \citet{yua2008} introduce a bicubic
$B$-splines basis.

One of the assumptions in the A98 DIA framework is that the photometric scaling between the reference image
and the target image is characterised by a single number, which may be a reasonable assumption for images
covering a small field-of-view (FOV), where spatial variations in atmospheric transparency and airmass are generally
negligible. However, DIA is now being applied in projects that generate images covering multiple square
degrees each (e.g. Palomar Transient Factory - \citealt{rau2009} and \citealt{law2009}, PanSTARRS - \citealt{kai2010}),
where non-uniform transparency is common (due to passing clouds)
and extinction varies from one edge of the image to another due to airmass gradients across the field.
Extension of the DIA framework to a spatially varying photometric scale factor is therefore a necessary
generalisation in the application of DIA to these projects.

In Section~\ref{sec:general_sol}, we take the step of generalising DIA to be able to cope with a spatially varying
photometric scale factor, while simultaneously modelling the spatial variation of the kernel shape and differential
background. In presenting this generalised formulation, we also take the opportunity to present a clear set of DIA
equations, with user-friendly notation, grouped in a logical way. The original DIA formulations in the literature (A98; A00)
are not so transparent in this respect, and the M08 formulation where delta basis functions are
introduced omits the consideration of pixel uncertainties, has difficult notation, and misses a number of
important simplifications with respect to this kernel basis (see Section~\ref{sec:delta_basis}). Discussion of the most
popular choices for the kernel basis functions and their implications with regard to the DIA formulation
is made in Section~\ref{sec:choice_for_basis}, where we also introduce the mixed-resolution delta basis functions.
In Section~\ref{sec:test}, we validate our algorithm using simulated data and we demonstrate it using some real data.
Section~\ref{sec:impl_and_opt} has been written to provide some implementation and optimisation hints for the DIA developer,
and the methodology that we propose will help to make the DIA algorithms more feasible with respect
to the increasing data volume (image sizes and numbers) from the latest generation of time-series imaging projects.
Finally, we state our conclusions in Section~\ref{sec:conclusions}.

\section{The General Difference Image Analysis Formulation And Solution}
\label{sec:general_sol}

In this Section, we derive a general theoretical formulation of the difference image analysis problem from which
all previously published formulations arise as special cases. This generalisation allows
us to exercise control separately over the spatial variation of each kernel basis function, the photometric scale factor,
and the differential sky background, as we show in Sections~\ref{sec:target_im_model}~\&~\ref{sec:spatial_var_ps}.

\subsection{Defining The Target Image Model}
\label{sec:target_im_model}

We start as in B08 by considering a pair of registered images sampled on the same pixel grid, one being
the reference image with pixel values $R_{ij}$, and the other the target image with pixel values $I_{ij}$, 
where $i$ and $j$ are pixel indices referring to the column $i$ and row $j$ of the image. We denote the
spatial coordinate system in these images by $(x,y)$, and the $(x,y)$ coordinates of the $(i,j)$th pixel
by $(x_{i},y_{j})$. Exact image registration is not strictly necessary, since the best formulations for
the kernel model include corrections for translational (but not rotational or otherwise) image misalignments,
which has the advantage of avoiding problematic image interpolation in many cases.

As first formulated by A98, we construct the model $M(x,y)$ for the target image as the reference image
convolved with a spatially varying kernel $K(u,v,x,y)$ (where $u$ and $v$ are kernel coordinates) plus
a spatially varying differential background $B(x,y)$:
\begin{equation}
M(x,y) = [R \otimes K](x,y) + B(x,y)
\label{eqn:model}
\end{equation}
We wish to determine the best-fit convolution kernel and differential background, and to do this we must first make
further assumptions about their functional form. We note that since the reference image is part of the target image
model, it may be desirable to also determine the reference image pixel values $R_{ij}$. However, finding a solution to this
issue is outside the scope of this paper.

A98 made the important step of decomposing the kernel into a set of basis functions thereby linearising the
expression in Equation~\ref{eqn:model}. Subsequently, A00 generalised the kernel decomposition to include the
spatial variation of the basis function coefficients, which facilitated the modelling of the spatial variation of the kernel.
We form the same kernel decomposition:
\begin{equation}
K(u,v,x,y) = \sum_{q = 1}^{N_{\kappa}} a_{q}(x,y) \, \kappa_{q}(u,v)
\label{eqn:ker_decomp1}
\end{equation}
where $\kappa_{q}(u,v)$ is the $q$th kernel basis function, $a_{q}(x,y)$ is the $q$th spatially variable coefficient,
and $N_{\kappa}$ is the number of kernel basis functions.

A polynomial is a sensible choice of model for the spatial variation of the kernel basis function coefficients since
it respects the linearity of the decomposition in Equation~\ref{eqn:ker_decomp1}, and by specifying the polynomial degree, one
may control the amount of spatial variation that is to be modelled. The polynomial form for $a_{q}(x,y)$ was adopted by
A00 with the same degree for each basis function coefficient. We generalise this further by modelling each coefficient
as a polynomial with individual degree $d_{q}$, providing a flexibility that we require later on:
\begin{equation}
a_{q}(x,y) = \sum_{m = 0}^{d_{q}} \sum_{n = 0}^{d_{q} - m} a_{qmn} \, \eta(x)^{m} \, \xi(y)^{n}
\label{eqn:ker_coeffs1}
\end{equation}
where the $a_{qmn}$ are polynomial coefficients for the $q$th kernel basis function. The coordinates $(\eta(x),\xi(y))$
are normalised spatial coordinates defined by:
\begin{eqnarray}
\eta(x) = (x - x_{c})/N_{x} \label{eqn:eta_x} \\
\xi(y) = (y - y_{c})/N_{y} \label{eqn:xi_y}
\end{eqnarray}
which follow from the Taylor expansion of the spatial coordinates $(x,y)$ around the image centre $(x_{c},y_{c})$ for an image
of size $N_{x} \times N_{y}$ pixels. This coordinate conversion improves the orthogonality of the spatial
polynomial terms\footnote{Although not considered here, further orthogonalisation of the spatial polynomial terms could be achieved by using,
for example, Gram-Schmidt orthogonalisation. However, the orthogonalisation can only ever be approximate
as the dot products that define orthogonality use inverse-variance pixel weights, and the variances depend
on the model being fitted (see Section~\ref{sec:uncert_and_iter}).}, and it
prevents the significant polynomial coefficients from becoming progressively smaller for the higher order polynomial terms.

As in A98, we also adopt a polynomial model of degree $d_{\mbox{\scriptsize B}}$ for the differential background:
\begin{equation} 
B(x,y) = \sum_{k = 0}^{d_{\mbox{\tiny B}}} \sum_{l = 0}^{d_{\mbox{\tiny B}} - k} b_{kl} \, \eta(x)^{k} \, \xi(y)^{l}
\label{eqn:back_coeffs1}
\end{equation}
where the $b_{kl}$ are the polynomial coefficients.

We now have a model $M(x,y)$ for the target image that is a linear combination of functions of $x$ and $y$.
This is easily shown by substituting Equations~\ref{eqn:ker_decomp1},~\ref{eqn:ker_coeffs1}~\&~\ref{eqn:back_coeffs1}
into Equation~\ref{eqn:model} and using the fact that convolution is distributive:
\begin{equation}
M(x,y) = \sum_{q = 1}^{N_{\kappa}} [R \otimes \kappa_{q}](x,y) \sum_{m = 0}^{d_{q}} \sum_{n = 0}^{d_{q} - m} a_{qmn} \, \eta(x)^{m} \, \xi(y)^{n}
         \, + \sum_{k = 0}^{d_{\mbox{\tiny B}}} \sum_{l = 0}^{d_{\mbox{\tiny B}} - k} b_{kl} \, \eta(x)^{k} \, \xi(y)^{l}
\label{eqn:model_lincomb}
\end{equation}

The target image is a discrete image of pixel values $I_{ij}$ and therefore we wish to evaluate
the model for the target image at the discrete pixel coordinates $(x_{i},y_{j})$. Let us use
$M_{ij}$ to represent the discrete model image $M(x_{i},y_{j})$ and $(\eta_{i},\xi_{j})$ to represent
the discrete coordinate array $(\eta(x_{i}),\xi(y_{j}))$. Then, using the fact that the
convolution of the reference image $R_{ij}$ with the continuous kernel basis function 
$\kappa_{q}(u,v)$ is equivalent to a discrete convolution (see Appendix~A), we have:
\begin{equation}
M_{ij} = \sum_{q = 1}^{N_{\kappa}} [R \otimes \kappa_{q}]_{ij} \sum_{m = 0}^{d_{q}} \sum_{n = 0}^{d_{q} - m} a_{qmn} \, \eta_{i}^{m} \, \xi_{j}^{n}
         \, + \sum_{k = 0}^{d_{\mbox{\tiny B}}} \sum_{l = 0}^{d_{\mbox{\tiny B}} - k} b_{kl} \, \eta_{i}^{k} \, \xi_{j}^{l}
\label{eqn:model_lincomb_discrete}
\end{equation}
with:
\begin{equation}
[R \otimes \kappa_{q}]_{ij} = \sum_{rs} R_{(i+r)(j+s)} \kappa_{qrs}
\label{eqn:discrete_conv}
\end{equation}
where $r$ and $s$ are pixel indices corresponding to the column $r$ and row $s$ of the discrete kernel
basis function $\kappa_{qrs}$ defined by:
\begin{equation}
\kappa_{qrs} = \int_{s - \frac{1}{2}}^{s + \frac{1}{2}} \int_{r - \frac{1}{2}}^{r + \frac{1}{2}} \! \kappa_{q}(u,v) \,\, \text{d}u \,\, \text{d}v
\label{eqn:discrete_ker_basis_func}
\end{equation}

We refer to $[R \otimes \kappa_{q}]_{ij}$ as a {\it basis image} since it is the linear combination
of these basis images modified by spatial polynomials and combined with the differential background that constitutes the target image model.
A basis image $[R \otimes \kappa_{q}]_{ij}$ is calculated from the discrete convolution of the reference
image $R_{ij}$ with the corresponding discrete kernel basis function $\kappa_{qrs}$ via Equation~\ref{eqn:discrete_conv},
which implies that the reference image $R_{ij}$ must extend beyond the pixel domain of the target image $I_{ij}$.
The discrete kernel basis function $\kappa_{qrs}$ may be defined
directly, or calculated by analytical or numerical integration of Equation~\ref{eqn:discrete_ker_basis_func}
given a definition for $\kappa_{q}(u,v)$. Note that the terms for modelling the differential background in
Equation~\ref{eqn:model_lincomb_discrete} can be thought of as multiplying a basis image that is set to unity at all
pixels.

All that is now required to fully define the model for the target image is to make a choice of suitable kernel
basis functions, from which the corresponding basis images are derived. This is where different authors have made
different choices (e.g. the Gaussian basis functions, the delta basis functions, etc.), and we leave the treatment
of these choices to Section~\ref{sec:choice_for_basis} where we consider their implications in more detail.

\subsection{The Kernel Model}
\label{sec:kernel_model}

Assuming that we have a solution for the polynomial coefficients $a_{qmn}$ of the kernel basis functions, we would
like to know how to construct the discrete kernel model $K_{rsij}$ at any pixel $(i,j)$ in the target image.
This is achieved by defining:
\begin{equation}
K_{rsij} = \int_{s - \frac{1}{2}}^{s + \frac{1}{2}} \int_{r - \frac{1}{2}}^{r + \frac{1}{2}} \! K(u,v,x_{i},y_{j})  \,\, \text{d}u \,\, \text{d}v
\label{eqn:define_discrete_kernel}
\end{equation}
which, on substitution of Equations~\ref{eqn:ker_decomp1},~\ref{eqn:ker_coeffs1}~\&~\ref{eqn:discrete_ker_basis_func},
reduces to:
\begin{equation}
K_{rsij} = \sum_{q = 1}^{N_{\kappa}} \kappa_{qrs} \sum_{m = 0}^{d_{q}} \sum_{n = 0}^{d_{q} - m} a_{qmn} \, \eta_{i}^{m} \, \xi_{j}^{n}
\label{eqn:kernel_model}
\end{equation}

\subsection{Controlling The Spatial Variation Of The Photometric Scale Factor}
\label{sec:spatial_var_ps}

The kernel sum $P_{ij} = \sum_{rs} K_{rsij}$, which in general is a function of spatial pixel $(i,j)$, defines the {\it photometric scale factor}
between the reference image and the target image:
\begin{equation}
P_{ij} = \sum_{rs} \sum_{q = 1}^{N_{\kappa}} \kappa_{qrs} \sum_{m = 0}^{d_{q}} \sum_{n = 0}^{d_{q} - m} a_{qmn} \, \eta_{i}^{m} \, \xi_{j}^{n}
\label{eqn:kernel_ps1}
\end{equation}

Our current formulation of the DIA problem in Section~\ref{sec:target_im_model} is such that $P_{ij}$ will vary across
the image as a polynomial of degree equal to the maximum of the set of degrees
$d_{\mbox{\scriptsize max}} = \max_{\,q} \left\{ d_{q} \right\}$ for the
coefficients of the (sub-)set of kernel basis functions that have a non-zero sum.
This can be seen by swapping the summation order in Equation~\ref{eqn:kernel_ps1} and combining the kernel basis function
coefficients into a single set of coefficients $a^{\,\prime}_{mn}$:
\begin{equation}
P_{ij} = \sum_{m = 0}^{d_{\mbox{\tiny max}}} \sum_{n = 0}^{d_{\mbox{\tiny max}} - m} a^{\,\prime}_{mn} \, \eta_{i}^{m} \, \xi_{j}^{n} \\
\label{eqn:kernel_ps2}
\end{equation}
where:
\begin{equation}
a^{\,\prime}_{mn} = \sum_{q = 1}^{N_{\kappa}} a_{qmn} \, \sum_{rs} \kappa_{qrs} 
\label{eqn:kernel_ps3}
\end{equation}

This behaviour may be undesirable if we wish to employ a different degree of spatial variation
in the photometric scale factor to the degree of spatial variation of the shape of the convolution kernel. A00
noted that those kernel basis functions with zero sums do not contribute to the spatial variation of the
photometric scale factor, regardless of the spatial variation of their coefficients, and that one may always
construct a new set of kernel basis functions that are a linear combination of the original set of basis functions.

We assume that our kernel basis functions have been normalised to a sum of unity, or have a zero sum, and that our 
first kernel basis function $\kappa_{1rs}$, without loss of generality, has a sum of unity.
We then form a new set of kernel basis functions as follows:
\begin{equation}
\kappa^{\,\prime}_{qrs} =
\begin{cases}
\kappa_{qrs}                &  \mbox{if $q = 1$ or $\,\sum_{rs} \kappa_{qrs} = 0$}   \\
\kappa_{qrs} - \kappa_{1rs} &  \mbox{if $q > 1$ and $\,\sum_{rs} \kappa_{qrs} = 1$}  \\
\end{cases}
\label{eqn:kernel_ps4}
\end{equation}
It follows that all of our new kernel basis functions $\kappa^{\,\prime}_{qrs}$ have zero sums except for the
first basis function $\kappa^{\,\prime}_{1rs}$ which has a sum of unity.

Adopting our new set of kernel basis functions and dropping the prime from our notation, the photometric scale factor $P_{ij}$ reduces to:
\begin{equation}
P_{ij} = \sum_{m = 0}^{d_{1}} \sum_{n = 0}^{d_{1} - m} a_{1mn} \, \eta_{i}^{m} \, \xi_{j}^{n}
\label{eqn:kernel_ps5}
\end{equation}
which is a polynomial in the spatial coordinates $(x,y)$ of degree $d_{1}$.

Hence, by transforming the kernel basis functions as outlined above, one may specify a polynomial degree $d_{1}$ of 
spatial variation for the photometric scale factor, associated only with the coefficient of the first kernel basis
function, and which we redefine as the degree $d_{\mbox{\scriptsize P}}$. Collectively, the spatial variation of the kernel basis functions
describes the kernel shape variations, and therefore the polynomial degree of spatial variation for the kernel shape
is set by the value of $\max_{\,q} \left\{ d_{q} \right\}$, which is always greater than or equal to $d_{\mbox{\scriptsize P}}$.
This is an important point to understand since if one wants to model the situation where the kernel shape is expected
to spatially vary with a smaller degree than the photometric scale factor, then one should still fit a model with
$\min_{\,q} \left\{ d_{q} \right\} = d_{\mbox{\scriptsize P}}$. For example, to model the situation where the kernel
shape is spatially invariant between two images but the spatial transparency pattern varies linearly (e.g. because
of changes in airmass gradient), then one must adopt a linear spatial variation for all of the kernel basis
functions. This enables the spatial variations of the zero-sum kernel basis functions to offset the spatial variations in kernel
shape induced by the spatial variations of the unit-sum kernel basis function.

To summarise, we have shown how to decouple the spatial variation of the photometric scale factor from the kernel shape
variations (with the aforementioned caveat), which leads to three natural types of spatial variation in the DIA formulation; namely, photometric scale
factor variations, differential background variations, and kernel shape variations, characterised by the degrees
$d_{\mbox{\scriptsize P}}$, $d_{\mbox{\scriptsize B}}$, and
$d_{\mbox{\scriptsize S}} = \max_{\,q} \left\{ d_{q} \right\} \ge d_{\mbox{\scriptsize P}}$, respectively.

\subsection{Fitting The Target Image Model}
\label{sec:fit_target_im_model}

In order to fit the model in Equation~\ref{eqn:model_lincomb_discrete} to the target image, we construct the chi-squared:
\begin{equation}
\chi^{2} = \sum_{ij} \left( \frac{I_{ij} - M_{ij}}{\sigma_{ij}} \right)^{2}
\label{eqn:chi_sqr}
\end{equation}
where the $\sigma_{ij}$ represent the target image pixel uncertainties.
Minimising the chi-squared in Equation~\ref{eqn:chi_sqr} falls into the class of general linear least-squares problems,
since the model in Equation~\ref{eqn:model_lincomb_discrete} is linear with respect to the unknown coefficients $a_{qmn}$ and $b_{kl}$
to be determined. This class of problems has a standard solution procedure by construction of the {\it normal equations}.
We refer the reader to the treatment of this subject in Numerical Recipes (\citealt{pre2007}) for more details.

The normal equations are most compactly represented by the matrix equation:
\begin{equation}
\mathbf{H} \boldsymbol{\alpha} = \boldsymbol{\beta}
\label{eqn:matrix_normal_eqns}
\end{equation}
where the square matrix $\mathbf{H}$ is the least-squares matrix, the vector $\boldsymbol{\alpha}$ is the vector of
model parameters, and $\boldsymbol{\beta}$ is another vector.

For each kernel basis function, there are \\ 
$N_{q} = (d_{q} + 1) (d_{q} + 2) / 2$ polynomial coefficients
$a_{qmn}$, and for the differential background, there are 
$N_{\mbox{\scriptsize B}} = (d_{\mbox{\scriptsize B}} + 1) (d_{\mbox{\scriptsize B}} + 2) / 2$ polynomial
coefficients $b_{kl}$, leading to a total of $N_{\mbox{\scriptsize par}} = (\sum_{q} N_{q}) + N_{\mbox{\scriptsize B}}$
parameters to be determined. Hence the least-squares matrix $\mathbf{H}$ is of size $N_{\mbox{\scriptsize par}}$ by
$N_{\mbox{\scriptsize par}}$ elements, and the vectors $\boldsymbol{\alpha}$ and $\boldsymbol{\beta}$ are of length $N_{\mbox{\scriptsize par}}$
elements.

If we take $z$ as a generalised index for all of the free parameters, then we are simply assigning a
one-to-one correspondence $f : z \leftrightarrow (q,m,n,k,l)$ that specifies which coefficient, $a_{qmn}$ or $b_{kl}$, 
corresponds to the current element $\alpha_{z}$ of the vector of parameters $\boldsymbol{\alpha}$.
This mapping may order the parameters in an arbitrary way, but the ordering is only important for the efficient
computation of $\mathbf{H}$ and $\boldsymbol{\beta}$ if one does not pre-calculate all of the necessary polynomial and
basis images (see Sections~\ref{sec:memory}~\&~\ref{sec:lsm}).

Following from the definition of the model for the target image in Equation~\ref{eqn:model_lincomb_discrete},
the elements of the least-squares matrix $\mathbf{H}$ (i.e. the coefficients
in the normal equations) and vector $\boldsymbol{\beta}$ may now be written out explicitly in terms
of the basis images:
\begin{equation}  
H_{z z^{\,\prime}} =
\begin{cases}
\sum_{ij} \, \eta^{m + m^{\,\prime}}_{i} \, \xi^{n + n^{\,\prime}}_{j} \, [R \otimes \kappa_{q}]_{ij} \; [R \otimes \kappa_{q^{\,\prime}}]_{ij} \, / \, \sigma^{2}_{ij} \\
\;\;\; \mbox{for $\alpha_{z} \equiv a_{qmn}$ and $\alpha_{z^{\,\prime}} \equiv a_{q^{\,\prime} m^{\,\prime} n^{\,\prime}}$} \\
\sum_{ij} \, \eta^{m + k^{\,\prime}}_{i} \, \xi^{n + l^{\,\prime}}_{j} \, [R \otimes \kappa_{q}]_{ij} \, / \, \sigma^{2}_{ij} \\
\;\;\; \mbox{for $\alpha_{z} \equiv a_{qmn}$ and $\alpha_{z^{\,\prime}} \equiv b_{k^{\,\prime} l^{\,\prime}}$} \\
\sum_{ij} \, \eta^{k + m^{\,\prime}}_{i} \, \xi^{l + n^{\,\prime}}_{j} \, [R \otimes \kappa_{q^{\,\prime}}]_{ij} \, / \, \sigma^{2}_{ij} \\  
\;\;\; \mbox{for $\alpha_{z} \equiv b_{kl}$ and $\alpha_{z^{\,\prime}} \equiv a_{q^{\,\prime} m^{\,\prime} n^{\,\prime}}$} \\
\sum_{ij} \, \eta^{k + k^{\,\prime}}_{i} \, \xi^{l + l^{\,\prime}}_{j} \, / \, \sigma^{2}_{ij} \\
\;\;\; \mbox{for $\alpha_{z} \equiv b_{kl}$ and $\alpha_{z^{\,\prime}} \equiv b_{k^{\,\prime} l^{\,\prime}}$} \\
\end{cases}
\label{eqn:least_squares_matrix}
\end{equation} 
\begin{equation}
\beta_{z} =
\begin{cases}
\sum_{ij} \, \eta^{m}_{i} \; \xi^{n}_{j} \; I_{ij} \; [R \otimes \kappa_{q}]_{ij} \, / \, \sigma^{2}_{ij} & \mbox{for $\alpha_{z} \equiv a_{qmn}$} \\
\sum_{ij} \, \eta^{k}_{i} \; \xi^{l}_{j} \; I_{ij} \; / \, \sigma^{2}_{ij} & \mbox{for $\alpha_{z} \equiv b_{kl}$} \\
\end{cases}
\label{eqn:least_squares_vector}
\end{equation}

Cholesky factorisation of the symmetric and positive-definite matrix $\mathbf{H}$, followed by forward and back substitution is
the most efficient and numerically stable method (\citealt{gol1996}) for obtaining the solution
$\boldsymbol{\alpha} = \boldsymbol{\widehat{\alpha}}$ to the normal equations. Explicit calculation of the matrix inverse
$\mathbf{H}^{-1}$ is only strictly necessary if one requires the covariance matrix
$cov \, (\widehat{\alpha}_{z},\widehat{\alpha}_{z^{\,\prime}}) = \left\{ \mathbf{H}^{-1} \right\}_{z z^{\,\prime}}$.
We note that the calculation of the uncertainties in the elements of $\boldsymbol{\widehat{\alpha}}$
is one such case since the uncertainty $\sigma_{z}$ in each $\widehat{\alpha}_{z}$ is given by:
\begin{equation}
\sigma_{z} = \sqrt{ \left\{ \mathbf{H}^{-1}  \right\}_{z z} }
\label{eqn:sol_uncertainties}
\end{equation}

\subsection{The Noise Model And Iteration}
\label{sec:uncert_and_iter}

The calculation of the least-squares matrix $\mathbf{H}$ and vector $\boldsymbol{\beta}$ requires the
adoption of a suitable noise model for the target image pixel uncertainties $\sigma_{ij}$. B08 specify
one such model as:
\begin{equation}
\sigma_{ij}^{2} = \frac{\sigma_{0}^{2}}{F_{ij}^{2}} + \frac{M_{ij}}{G \, F_{ij}}
\label{eqn:noise_model}
\end{equation}
where $\sigma_{0}$ is the CCD readout noise (ADU), $G$ is the CCD gain (e$^{-}$/ADU), and $F_{ij}$ is the
master flat-field image. This model assumes that both the master flat-field image $F_{ij}$ and the reference image 
$R_{ij}$ are noiseless, which is a reasonable assumption for such typically high signal-to-noise (S/N) images.

Most importantly, we note that in this noise model, the uncertainties $\sigma_{ij}$ depend on the target
image model $M_{ij}$ and consequently, fitting $M_{ij}$ as described in Section~\ref{sec:fit_target_im_model}
becomes an iterative process\footnote{Strictly speaking, the fact that the uncertainties $\sigma_{ij}$ depend on the
target image model $M_{ij}$ also implies that minimising $\chi^{2}$ is no longer equivalent to maximising the likelihood.
The maximum likelihood estimator is obtained instead by minimising $\chi^{2} + \sum_{ij} \ln ( \sigma_{ij}^{2} )$,
which renders the fitting of the target image model as a non-linear problem.}.
In the first iteration, it is appropriate to approximate $M_{ij}$ by 
using $I_{ij}$, which enables the calculation of the initial kernel and differential background solution.
In subsequent iterations, the current image model defined by Equation~\ref{eqn:model_lincomb_discrete}
should be used to set the $\sigma_{ij}$ as per Equation~\ref{eqn:noise_model}. In Appendix~B, we 
use an example to demonstrate the bias that can be introduced into the model parameters if the iterative
fitting procedure is not performed (see also Section~\ref{sec:simdata}).

It is also desirable to employ a $k$-sigma-clip algorithm
in order to prevent outlier target image pixel values from influencing the solution, including
those from variable objects and cosmic ray events. This may easily be
achieved by calculating the normalised residuals $\epsilon_{ij} = (I_{ij} - M_{ij}) / \sigma_{ij}$
and ignoring any pixels with $\left| \epsilon_{ij} \right| \ge k$ in subsequent iterations. The reliability of the 
$k$-sigma-clip algorithm depends heavily on the accuracy of the adopted noise model, and since
the initial $\sigma_{ij}$ values are calculated using an approximation to $M_{ij}$, we recommend that
the sigma-clipping commences at the second iteration. 

Our final note in this Section is that the noise model in Equation~\ref{eqn:noise_model} could be
improved, specifically by considering the noise introduced by the reference image, which is non-negligible
when the S/N of the reference image is similar to that of the target image. A00 and B08 have
previously considered such a noise model. Here, we explicit
a useful noise model for a target image and a combined reference image that have been registered to the nearest pixel
(i.e. avoiding image resampling):
\begin{equation}
\sigma_{ij}^{2} = \frac{\sigma_{0}^{2}}{F_{\mbox{\scriptsize tar},ij}^{2}} + \frac{M_{ij}}{G \, F_{\mbox{\scriptsize tar},ij}}
                + \sum_{rs} K_{rsij}^{2} \, \sigma_{\mbox{\scriptsize ref},(i+r)(j+s)}^{2}
\label{eqn:noise_model_incl_ref}
\end{equation}
with:
\begin{equation}
\sigma_{\mbox{\scriptsize ref},ij}^{2} = \frac{1}{N_{\mbox{\scriptsize im}}^{2}} \sum_{k}
                                         \left[ \frac{\sigma_{0}^{2}}{F_{\mbox{\scriptsize ref},kij}^{2}}
                                         + \frac{R_{kij}^{\,\prime}}{G \, F_{\mbox{\scriptsize ref},kij}} \right]
\label{eqn:noise_model_ref}
\end{equation}
where the $R_{kij}^{\,\prime}$ represent the $N_{\mbox{\scriptsize im}}$ images that have been combined to create the reference image,
and $F_{\mbox{\scriptsize tar},ij}$ and $F_{\mbox{\scriptsize ref},kij}$ are the master flat-field images corresponding to the
target image and constituent images of the reference image, respectively.

\subsection{The Input Data}
\label{sec:input}

Ideally, {\it every pixel} in the target image should be used
in the calculation of $\mathbf{H}$ and $\boldsymbol{\beta}$, and therefore contribute to the kernel and differential background
solution. However, due to the nature of the convolution process, the target image model is undefined in a border of width
equal to half the kernel width around the image edges if the reference image is the same size as the target image,
and therefore these target image pixels cannot be used in the
calculation of $\mathbf{H}$ and $\boldsymbol{\beta}$. Also, ``bad'' pixels (e.g. bad columns/rows, hot pixels, saturated pixels,
cosmic-ray events, etc.) should be excluded from the calculations, which
means that any target image pixel $(i,j)$ to be included in the calculation of $\mathbf{H}$ and $\boldsymbol{\beta}$ should
be ``good'' in the target image, and that all reference image pixels to be used for calculating the target image model at $(i,j)$
should be ``good'' in the reference image. This implies that a bad pixel in the reference image can discount a set
of pixels equal to the kernel area in the target image, and therefore, as suggested in B08, bad pixels in the reference image
should be kept to a minimum, and kernels with excessively large footprints should be avoided when there are bad pixels in the
reference image (e.g. see Section~2.3 of \citealt{bra2011}).

The areas of the target image which contain only sky background and no astronomical objects will only contribute 
information on the differential sky background coefficients in the target image model. Hence, one may limit the set
of target image pixels to be used in the calculation of $\mathbf{H}$ and $\boldsymbol{\beta}$ to a set of image sub-regions
encompassing the higher S/N objects in the target image, which speeds the computations (fewer pixel values to
be included in the required summations) while sacrificing some information. We note that contrary to the statements of
some authors (e.g. M08), these sub-regions need not be centred on isolated stars. In fact, sub-regions of crowded
high S/N objects (PSF-like or not) are precisely the image regions that contain the most information on the convolution kernel
and differential background, because each pixel contains PSF and background information at a high S/N ratio.

\subsection{Difference Images}
\label{sec:diffim}

We briefly mention that the definition of a difference image $D_{ij}$ is:
\begin{equation}
D_{ij} = I_{ij} - M_{ij}
\label{eqn:diff_image}
\end{equation}
This image of residuals consists of noise (mainly Poisson noise from photon counting) and any differential
flux from objects that have varied in brightness and/or position compared to the epoch of the reference image, since
constant sources are fully subtracted during the DIA process. However, if an inappropriate kernel and/or differential
background model is chosen, then unwanted systematic errors will leave signatures in the difference image
as large-amplitude high-spatial-frequency residuals at the positions of the brighter
objects (for inappropriate kernel models), and as lower-amplitude low-spatial-frequency deviations in the difference
image background from zero (for inappropriate differential background models).
We note that if a reliable noise model exists, then the normalised difference image $\epsilon_{ij}$ defined by:
\begin{equation}
\epsilon_{ij} = \frac{I_{ij} - M_{ij}}{\sigma_{ij}}
\label{eqn:diff_image_norm}
\end{equation}
acts as a useful guide to the level of flux variation in any one pixel, since the pixel values in this image are in units of
sigma-deviations.

The purpose of producing a difference image is to enable accurate differential photometry to be performed in the absence of PSF crowding
for all objects of interest (constant and variable). The object positions are presumed known from analysis
of the reference image or from fitting of the differential flux on the difference image.

\section{Common Basis Function Choices}
\label{sec:choice_for_basis}

In this Section, we elucidate the common choices for the kernel basis functions. We stress that since the choice of
basis functions is fully independent of the DIA framework presented in the previous Section, the generation of a
set of basis functions may be implemented as code that is completely separate from the DIA code.

\subsection{The Gaussian Basis Functions}
\label{sec:gauss_basis}

A98 introduced the {\it Gaussian basis functions} as a set of two-dimensional radially-symmetric Gaussian functions
of different widths, each one modified by a polynomial of the kernel coordinates of a certain degree. The justifications
for this choice are that an instrumental PSF is approximated by a Gaussian to first order, the convolution of a Gaussian
by a Gaussian is also a Gaussian, and that a Gaussian decays rapidly beyond a given distance. The user is required to
specify the number of Gaussian functions $N_{\mbox{\scriptsize gau}}$, and then for each Gaussian function (indexed by
$\lambda$), the user must specify the width $\sigma_{\mbox{\scriptsize gau},\lambda}$ and the degree of the modifying polynomial
$D_{\mbox{\scriptsize gau},\lambda}$. It follows that the definition of the $q$th kernel basis function corresponding to the $\lambda$th Gaussian
with a modifying polynomial term of degree $d_{\mbox{\scriptsize gau},u}$ and degree $d_{\mbox{\scriptsize gau},v}$ in
the $u$ and $v$ coordinates, respectively, is given by:
\begin{equation}
\kappa_{q}(u,v) = u^{\,d_{\mbox{\tiny gau},u}} \, v^{\,d_{\mbox{\tiny gau},v}} \, e^{-(u^{2} + v^{2}) / 2 \sigma_{\mbox{\tiny gau},\lambda}^{2}}
\label{eqn:gauss_basis}
\end{equation}
where $0 \le d_{\mbox{\scriptsize gau},u} + d_{\mbox{\scriptsize gau},v} \le D_{\mbox{\scriptsize gau},\lambda}$.
The number of kernel basis functions $N_{\kappa}$ in this prescription is given by:
\begin{equation}
N_{\kappa} = \sum_{\lambda = 1}^{N_{\mbox{\tiny gau}}} \frac{(D_{\mbox{\scriptsize gau},\lambda} + 1)(D_{\mbox{\scriptsize gau},\lambda} + 2)}{2}
\label{eqn:num_gauss_basis}
\end{equation}

The Gaussian basis functions need to be numerically integrated via Equation~\ref{eqn:discrete_ker_basis_func}
to form the corresponding discrete kernel basis functions, and then subsequently they
should be transformed as detailed in Section~\ref{sec:spatial_var_ps} to allow
control over the spatial variation of the photometric scale factor. Finally, we note that the
adoption of a set of Gaussian kernel basis functions does not provide any simplification in the calculation
of the basis images $[R \otimes \kappa_{q}]_{ij}$ via Equation~\ref{eqn:discrete_conv}.

Typical specifications for the Gaussian basis functions in the literature usually include three Gaussian functions,
and the {\tt ISIS2.2}\footnote{http://www2.iap.fr/users/alard/package.html} software developed by A98 and A00 adopts
Gaussian widths of 0.7, 2.0, and 4.0~pix with modifying polynomials of degrees 6, 4, and 3, respectively, by default, resulting
in 53 Gaussian basis functions.
\citet{isr2007} investigated how the optimal choice of Gaussian basis functions depends on the properties
of the images for which DIA is to be performed (e.g. seeing, S/N, etc.), and although they manage to give
some general recommendations, there seems to be no unique answer. It has also been noted by \citet{yua2008}
that the radial symmetry of the Gaussian functions may not be appropriate for elliptical PSFs, although it would
be trivial to expand the Gaussian basis function definition in Equation~\ref{eqn:gauss_basis} to include elliptical
two-dimensional Gaussians with an arbitrary centre and axis orientation.

\subsection{The Delta Basis Functions}
\label{sec:delta_basis}

Let us introduce the definition of the Kronecker delta-function $\delta_{ij}$:
\begin{equation}
\delta_{ij} =   
\begin{cases}
1 & \mbox{if $i=j$} \\
0 & \mbox{if $i \ne j$} \\
\end{cases}
\label{eqn:kronecker_delta}
\end{equation}
Let us also assume that there exists a one-to-one correspondence $g : q \leftrightarrow (\mu,\nu)$ which associates the $q$th kernel basis function
with the discrete kernel pixel coordinates $(\mu,\nu)$ such that, without loss of generality, 
$q = 1 \Leftrightarrow (\mu,\nu) = (0,0)$. Then we may directly define the $q$th discrete kernel basis function $\kappa_{qrs}$ by:
\begin{equation}
\kappa_{qrs} =
\begin{cases}
\delta_{r0} \, \delta_{s0}                                  & \mbox{for $q = 1$}   \\
\delta_{r\mu} \, \delta_{s\nu} - \delta_{r0} \, \delta_{s0} & \mbox{for $q > 1$} \\
\end{cases}
\label{eqn:delta_basis}
\end{equation}
where we have already included the transformation as detailed in Section~\ref{sec:spatial_var_ps} to allow
control over the spatial variation of the photometric scale factor.
It is clear that when $q = 1$, $\kappa_{1rs}$ obtains the value of 1 at $(r,s) = (0,0)$ and 0 elsewhere,
and that when $q > 1$, $\kappa_{qrs}$ obtains the value of 1 at $(r,s) = (\mu,\nu)$, $-$1 at $(r,s) = (0,0)$, and 0 elsewhere.
Thus $\kappa_{1rs}$ adds flux to the PSF core, and the other $\kappa_{qrs}$ subtract flux from the core and add it back at displaced locations.

We refer to this set of kernel basis functions as the
{\it delta basis functions}. The set of delta basis functions may be chosen to cover any discrete
kernel domain (e.g. circular - B08, square - M08, etc.) by defining the number of kernel basis functions $N_{\kappa}$ and the mapping
$g$ appropriately.

The basis images corresponding to the delta basis functions have a conveniently simple form that may
be derived by substituting Equation~\ref{eqn:delta_basis} into Equation~\ref{eqn:discrete_conv} and including
a product of delta-functions to combine the two cases into one expression:
\begin{equation}
[R \otimes \kappa_{q}]_{ij} = R_{(i+\mu)(j+\nu)} + (\delta_{\mu 0} \, \delta_{\nu 0} - 1) \, R_{ij}
\label{eqn:delta_basis_images}
\end{equation}
Hence, the first basis image is the reference image itself, and the remaining basis images are
each formed by shifting the reference image by the appropriate integer-pixel shift, and then
subtracting the non-shifted reference image. This has important speed and memory implications when
implementing the calculation of the least-squares matrix and vector (see Section~\ref{sec:impl_and_opt}).

B08 introduced the idea of solving directly for the kernel pixel values $K_{rs}$ of a spatially invariant kernel.
We note that if we take $\kappa_{qrs} = \delta_{r\mu} \, \delta_{s\nu}$ for all $q$ as an alternative definition for the
discrete kernel basis functions in Equation~\ref{eqn:delta_basis}, then the corresponding basis images are given by
$[R \otimes \kappa_{q}]_{ij} = R_{(i+\mu)(j+\nu)}$.
This definition ignores any control that we may wish to exercise over the photometric scale factor, but this is not an issue
when considering a spatially invariant kernel (as in B08). Substitution of this new result for the basis images into
Equations~\ref{eqn:least_squares_matrix}~\&~\ref{eqn:least_squares_vector}, and assuming that the kernel and
differential background are spatially invariant, leads directly to the least-squares matrix and vector derived by
B08 from their direct solution approach. Hence, adoption of the delta basis functions in the A98 DIA framework
is equivalent to solving directly for the kernel pixel values. A similar line of argument extends this conclusion to
spatially variable kernels.

The delta basis functions require minimal information from the user about the kernel shape and size for their specification.
However, the dependence of the optimal kernel shape and size on the reference and target image properties has not yet been investigated,
although it is clear that the greater the difference in PSF width between the images, the larger the size of the convolution
kernel that is required to match the PSFs.

\subsection{The Mixed-Resolution Delta Basis Functions}
\label{sec:delta_basis_mix}

The number of delta basis functions, and hence the number of coefficients $a_{qmn}$, grows as the
number of kernel pixels, which in turn grows as the square of the kernel radius. Since the least-squares matrix
is a square matrix of size $N_{\mbox{\scriptsize par}}$ by $N_{\mbox{\scriptsize par}}$ elements, the number
of elements to be calculated in the least-squares matrix grows as the kernel radius to the fourth power.
Hence, the time taken to calculate the solution for the coefficients $a_{qmn}$ and $b_{kl}$ increases
considerably when solving for larger kernels.

\begin{figure}
\centering
\epsfig{file=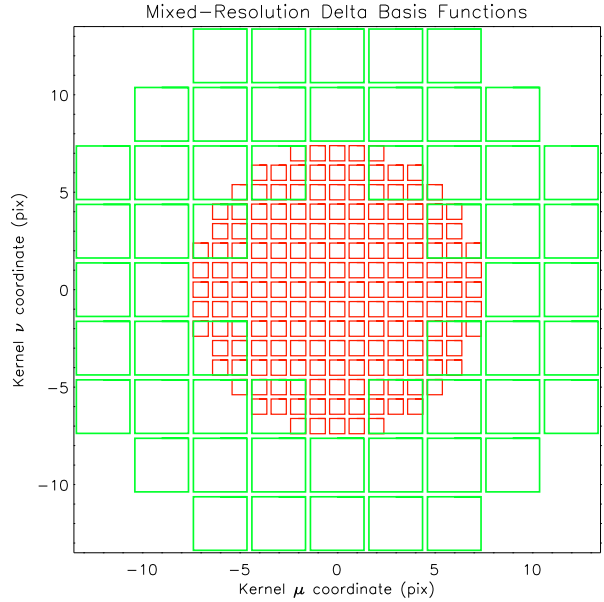,angle=0.0,width=\linewidth}
\caption{The distribution of single (red) and 3$\times$3 binned (green) kernel pixels for a circular kernel of radius
         13~pix that uses 3$\times$3 binned kernel pixels beyond a radius of 7~pix.
         \label{fig:ker_pix}}
\end{figure}

To address this performance issue, \citet{alb2009} introduced the idea of ``binned'' kernel pixels in the
outer part of the kernel on the assumption that the kernel shows slower variations of smaller amplitude beyond
a certain radius. Specifically, they introduced 3$\times$3 binned kernel pixels beyond a kernel radius of 7~pix to replace
the single kernel pixels, which greatly reduces the number of parameters to be solved for
while maintaining a sufficiently large kernel. For example, for a circular kernel of radius 13~pix, which fits
in a square array of 27 by 27 pixels, there are 577 single kernel pixels. Adopting the 3$\times$3
binned kernel pixels beyond a radius of 7~pix results in 233 parameters, of which 177 are single
kernel pixels and 56 are 3$\times$3 binned kernel pixels. The number of elements to be calculated in the least-squares matrix
is consequently reduced to $\sim$16\% without compromising the extent of the kernel model.
Figure~\ref{fig:ker_pix} shows the distribution of single (red) and 3$\times$3 binned (green) kernel pixels for this example.

We generalise the idea of a binned kernel pixel to that of an {\it extended delta basis function} defined by:
\begin{equation}
\kappa_{qrs} =
\begin{cases}
(1 / N_{\mbox{\scriptsize pix},q}) + (\delta_{\mu 0} \, \delta_{\nu 0} - 1) \, \delta_{r 0} \, \delta_{s 0} & \mbox{for $(r,s) \in S_{q}$}    \\
(\delta_{\mu 0} \, \delta_{\nu 0} - 1) \, \delta_{r 0} \, \delta_{s 0}                                      & \mbox{for $(r,s) \notin S_{q}$} \\
\end{cases}
\label{eqn:delta_basis_binned}
\end{equation}
where $S_{q}$ is the set of kernel pixels spanned by the extended delta basis function
(of any shape and spatial distribution), and
$N_{\mbox{\scriptsize pix},q}$ is the number of elements in $S_{q}$. Note that we have assumed that the
one-to-one correspondence $g : q \leftrightarrow (\mu,\nu)$ is defined with $q = 1 \Leftrightarrow (\mu,\nu) = (0,0)$.
Again, we have forced the sum of the extended
delta basis function to be unity if it is the first kernel basis function ($q=1$), and to be zero if it is not ($q > 1$),
in order to be able to exercise control over the spatial variation of the photometric scale factor.

The basis image corresponding to the extended delta basis function defined in Equation~\ref{eqn:delta_basis_binned}
is easily derived by substitution into Equation~\ref{eqn:discrete_conv}:
\begin{equation}
[R \otimes \kappa_{q}]_{ij} = \frac{1}{N_{\mbox{\scriptsize pix},q}} \left[ \sum_{(r,s) \in S_{q}} R_{(i+r)(j+s)} \right]
                            + (\delta_{\mu 0} \, \delta_{\nu 0} - 1) \, R_{ij}
\label{eqn:delta_basis_binned_images}
\end{equation}
Therefore, this basis image is formed by averaging
$N_{\mbox{\scriptsize pix},q}$ versions of the reference image with each one shifted by the appropriate
integer-pixel shift, and then subtracting the non-shifted reference image (except if this is the first
basis image). Again, this has important speed and memory implications when implementing the calculation of the
least-squares matrix and vector (see Section~\ref{sec:impl_and_opt}). We note that the basis images corresponding to the 3$\times$3
binned kernel pixels from \citet{alb2009} are formed from integer-pixel shifted versions of a box-car
smoothed reference image.

We refer to a set of basis functions as {\it mixed-resolution} if they include any combination of
delta basis functions and extended delta basis functions, and we emphasise that extended delta basis functions
need not be square and they may be of any shape (e.g. circles, rectangles, rings, arcs, etc.). We also note
that the delta basis function is a special case of the extended delta basis function, and that overlapping extended
delta basis functions are acceptable in a set of kernel basis functions as long as none of the extended delta basis functions
may be constructed as a linear combination of any of the other kernel basis functions. If this condition is not met, then the solution
for the coefficients in the target image model is degenerate. Finally, we mention that mixed-resolution delta basis
functions have the potential to be used in kernels with an adaptive resolution, which is a subject that has not yet been
investigated in terms of its application to DIA.

\section{Validating And Demonstrating The Algorithm}
\label{sec:test}

So far we have only examined the theory of our general DIA formulation. We now proceed to validate the
algorithm using simulated images. We also demonstrate the ability of the algorithm to correct for
a spatially varying differential transparency across the image area using real data.

\subsection{Simulated Image Data}
\label{sec:simdata}

Our first task is to check that the algorithm can recover the exact model coefficients used to generate a set
of simulated image data without any artificial noise added to the pixel values. By doing this we are
validating our DIA formulation by confirming that there are no degeneracies in the target image model that
we did not foresee.

We generate a reference image of size 1000$\times$1000~pix with a constant sky level of 1000~ADU and
with 5000 stars. The stars are generated using a Gaussian PSF with a full-width half-maximum (FWHM) of
4~pix, pixel coordinates drawn from a uniform distribution over the detector area, and log-fluxes drawn
from a uniform distribution between 2 and 5 (i.e. stars have fluxes between 10$^{2}$ and 10$^{5}$~ADU).
The image parameters that we have chosen are actually not important, and the tests in the absence of
artificial noise give the same results so long as there are at least a few stars spread out over the image.

We then generate a set of target images from the reference image using Equation~\ref{eqn:model_lincomb_discrete}
for various sets of kernel basis functions (Gaussian, delta, and mixed-resolution) and values for the corresponding
coefficients, and for all combinations of $d_{\mbox{\scriptsize P}}$, $d_{\mbox{\scriptsize B}}$, and
$d_{\mbox{\scriptsize S}}$ (defined in Section~\ref{sec:spatial_var_ps})
taken from the set $\left\{ 0,1,2,3 \right\}$. We find that when we fit each target image with the
model used to generate it, we can recover the exact input values (to within numerical precision) of the coefficients $a_{qmn}$ and $b_{kl}$
in Equation~\ref{eqn:model_lincomb_discrete} for all cases. Hence we confirm that our algorithm works and that
there are no hidden degeneracies.

Next we generate a set of target images from the reference image by convolving the reference image with a spatially
varying kernel of polynomial degree $d^{\,\prime}_{\mbox{\scriptsize S}}$ with the kernel normalised to a unit sum at each pixel.
Then we multiply the convolved reference image with a polynomial surface of degree $d^{\,\prime}_{\mbox{\scriptsize P}}$
representing the photometric scale factor and we add a polynomial surface of degree $d^{\,\prime}_{\mbox{\scriptsize B}}$
representing the differential background. We have done this for all combinations of $d^{\,\prime}_{\mbox{\scriptsize P}}$,
$d^{\,\prime}_{\mbox{\scriptsize B}}$, and $d^{\,\prime}_{\mbox{\scriptsize S}}$ taken from the set $\left\{ 0,1,2 \right\}$.
In this set up, the degree of spatial
variation of the kernel shape is actually $d^{\,\prime}_{\mbox{\scriptsize P}} + d^{\,\prime}_{\mbox{\scriptsize S}}$ since the polynomial surface
for the photometric scale factor multiplies the kernel pixel values which also spatially vary as a polynomial. Therefore, the
appropriate (linear) target image model has $d_{\mbox{\scriptsize P}} = d^{\,\prime}_{\mbox{\scriptsize P}}$,
$d_{\mbox{\scriptsize B}} = d^{\,\prime}_{\mbox{\scriptsize B}}$, and
$d_{\mbox{\scriptsize S}} = d^{\,\prime}_{\mbox{\scriptsize P}} + d^{\,\prime}_{\mbox{\scriptsize S}}$,
and when we adopt such a model we find that we can recover the exact values for the model coefficients (again to within numerical precision).
If we naively set $d_{\mbox{\scriptsize P}} = d^{\,\prime}_{\mbox{\scriptsize P}}$, $d_{\mbox{\scriptsize B}} = d^{\,\prime}_{\mbox{\scriptsize B}}$,
and $d_{\mbox{\scriptsize S}} = d^{\,\prime}_{\mbox{\scriptsize S}}$ for our target image model, then the algorithm does not
manage to perfectly fit the target image, leaving significant residuals.

Now we consider how the algorithm performs for simulated images with added artificial noise.
We adopt the same reference image as before and we use delta basis functions with
$d_{1} = d_{\mbox{\scriptsize P}} = 1$ and $d_{q} = d_{\mbox{\scriptsize S}} = 2$ for all $q > 1$. We define the kernel model to be a square array
of 7$\times$7 pixels. The target image model coefficients are arbitrarily chosen and specifically we set
$a_{1mn} = \left\{ 1.1, 0.3, 0.1 \right\}$ for $(m,n) = \left\{ (0,0), (1,0), (0,1) \right\}$. Also, we define
$d_{\mbox{\scriptsize B}} = 0$ and set $b_{00} = 100$. We then use all of these definitions in Equation~\ref{eqn:model_lincomb_discrete}
to generate a noiseless target image $S_{ij}$.

From the noiseless target image $S_{ij}$, we generate 10$^{3}$ noisy versions. Each noisy target image $I_{ij}$ is formed
by generating a 1000$\times$1000 pixel image $\Sigma_{ij}$ of values drawn from a normal distribution with zero mean and unit $\sigma$,
and then computing:
\begin{equation}
I_{ij} = S_{ij} + \Sigma_{ij} \sqrt{ \sigma_{0}^{2} + S_{ij} }
\end{equation}
where the coefficient of $\Sigma_{ij}$ is derived from Equation~\ref{eqn:noise_model} for
$G = 1$~e$^{-}$/ADU and $F_{ij} = 1$. We adopt a reasonable value for the readout noise of $\sigma_{0} = 5$~ADU.
For each noisy target image, we fit the same model used to generate the noiseless target image, employing the iterative
scheme described in Section~\ref{sec:uncert_and_iter} (but without sigma-clipping).

\begin{figure*}
\centering
\epsfig{file=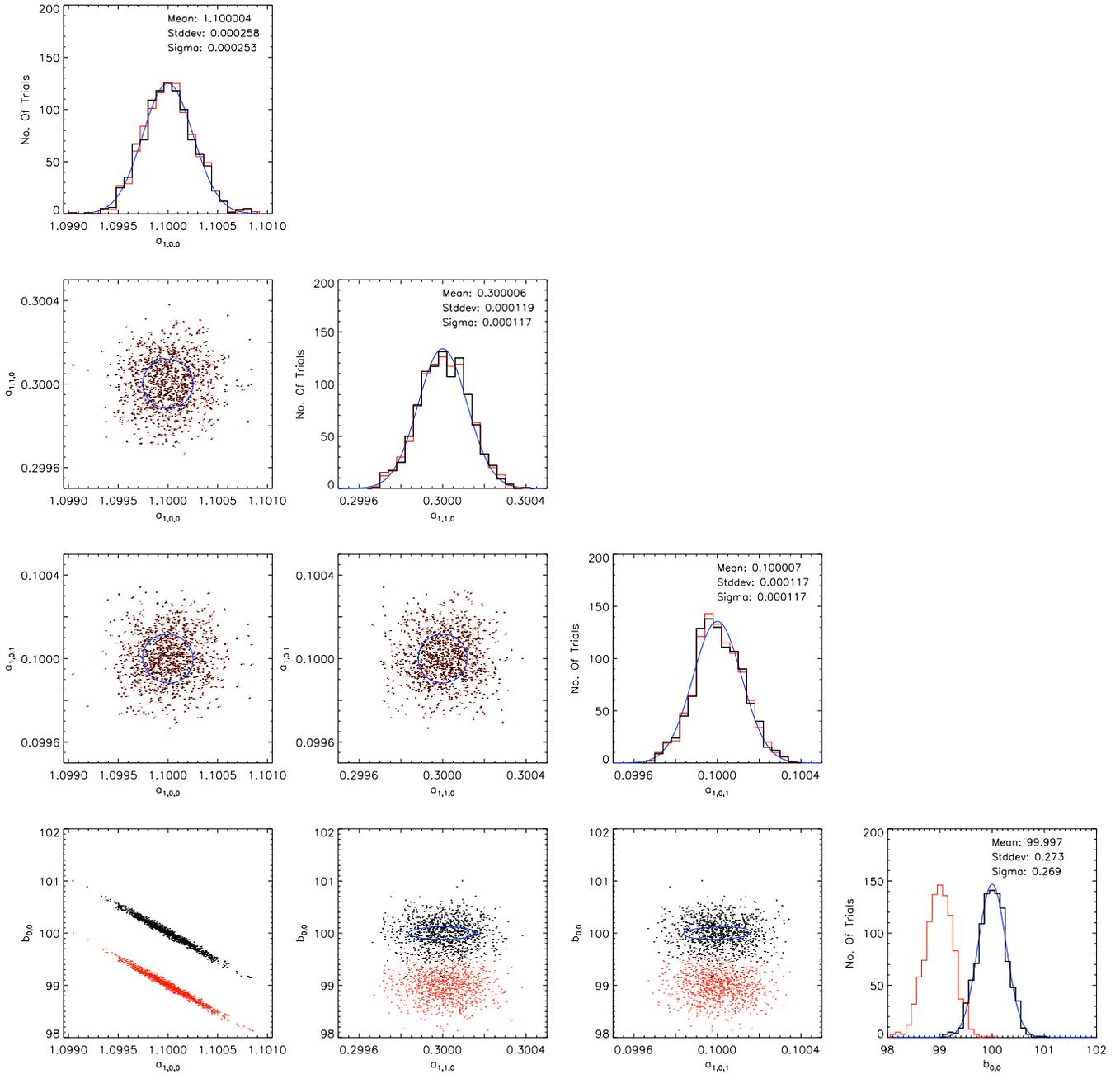,angle=0.0,width=\linewidth}
\caption{{\bf Plots along the diagonal:}
         Histograms of the coefficients $a_{1mn}$ for $(m,n) = \left\{ (0,0), (1,0), (0,1) \right\}$ and $b_{00}$
         as derived from the fits to the 10$^{3}$ noisy target images. The red and black histograms represent the coefficient distributions after the first
         and third iterations, respectively. The blue curves are fitted Gaussian distributions centred on
         the input coefficient values and with widths equal to the formal uncertainties in the coefficients.
         {\bf Off-diagonal plots:}
         Scatter plots for all of the coefficient pairs that can be formed from $a_{100}$, $a_{110}$, $a_{101}$, and $b_{00}$ using
         the results of the fits to the 10$^{3}$ noisy target images. The red and black points represent the fitted coefficients
         after the first and third iterations, respectively. The blue curves are formal 1$\sigma$-error ellipses.
         \label{fig:parhist}}
\end{figure*}

In the plots along the diagonal of Figure~\ref{fig:parhist}, we show the distributions of the coefficients
$a_{1mn}$ for $(m,n) = \left\{ (0,0), (1,0), (0,1) \right\}$ and $b_{00}$
as derived from the fits to the 10$^{3}$ noisy target images. The red and black histograms represent the coefficient distributions after the first
and third iterations, respectively, and the reason for iterating the solution is clear; namely, approximating $M_{ij}$ with $I_{ij}$ in
the noise model in Equation~\ref{eqn:noise_model} in the first iteration introduces a significant bias into the fitted coefficients (in this example $b_{00}$
is underestimated by $\sim$1~ADU or $\sim$1\%; see also Appendix~B). We also report in the plots the measured mean and standard deviation of each
coefficient distribution after the third iteration. The measured means of the coefficient distributions are an excellent match to
the input coefficient values (no differences to at least 5 significant figures), and the measured standard deviations are an excellent match to the formal
uncertainties in the coefficients reported by the algorithm
(calculated via Equation~\ref{eqn:sol_uncertainties} and displayed as ``Sigma'').
For the coefficient distributions after the third iteration,
we fit a Gaussian with mean and sigma equal to the corresponding input coefficient value and the formal uncertainty in the
coefficient, respectively, and we plot the Gaussian fits as the blue curves. One can see that the
coefficient distributions follow the Gaussian distributions very well.

In the off-diagonal plots of Figure~\ref{fig:parhist}, we show scatter plots for all of the coefficient pairs that can be formed
from $a_{100}$, $a_{110}$, $a_{101}$, and $b_{00}$ using the results of the fits to the 10$^{3}$ noisy target images. The red and black
points represent the fitted coefficients after the first and third iterations, respectively. In each plot we also display the
formal 1$\sigma$-error ellipses (blue curves) as provided by the covariance matrix of the fit (see Section~\ref{sec:fit_target_im_model}).
It is encouraging to see that there are virtually no correlations between the target image model coefficients $a_{100}$, $a_{110}$,
and $a_{101}$ associated with the spatial variation of the photometric scale factor, or between the differential background
coefficient $b_{00}$ and $a_{110}$ or $a_{101}$. Also, as expected, there is a strong anti-correlation between the zeroth-order
coefficients for the photometric scale factor and the differential background, $a_{100}$ and $b_{00}$.

The anti-correlation between $a_{100}$ and $b_{00}$ is a well-known feature of DIA that occurs when the reference image
includes a non-zero background level. The kernel basis functions with non-zero sums in the target image model
(Equation~\ref{eqn:model_lincomb_discrete}) serve to blur {\it and scale} the reference image, including the background level,
and hence the model terms for the differential background must compensate for this effect in the opposite sense. The consequence is that
if the photometric scale is overestimated, then the differential background will be underestimated to compensate,
and vice versa. To minimise the amplitude of this anti-correlation, we recommend subtracting the sky background from the
reference image before applying DIA (as suggested by B08 in their Section~2.2), a procedure which could also include the subtraction
of the spatially varying components of the background (i.e. to flatten the background).

The results of our investigations in this Section lead us to conclude that our DIA algorithm is working exactly as
expected for simulated images with added artificial noise.

\subsection{Real Image Data}
\label{sec:realdata}

We demonstrate our DIA algorithm using a pair of calibrated images from a commercial telescope (Celestron 8-inch Schmidt-Cassegrain
$f = 2032$~mm) and CCD camera (Kodak KAF-1603ME) with a pixel scale of $\sim$1.8~arcsec/pix, a FOV of 0.26$\times$0.26~degrees, and no filter.
We designate one of the images as the reference image and the other as the target image. Since both images are undersampled, it is
necessary to pre-blur them before applying DIA.
We blur the reference and target images with Gaussian convolution kernels of FWHMs 3.5 and 4.0~pix, respectively.

\begin{figure*}
\centering
\epsfig{file=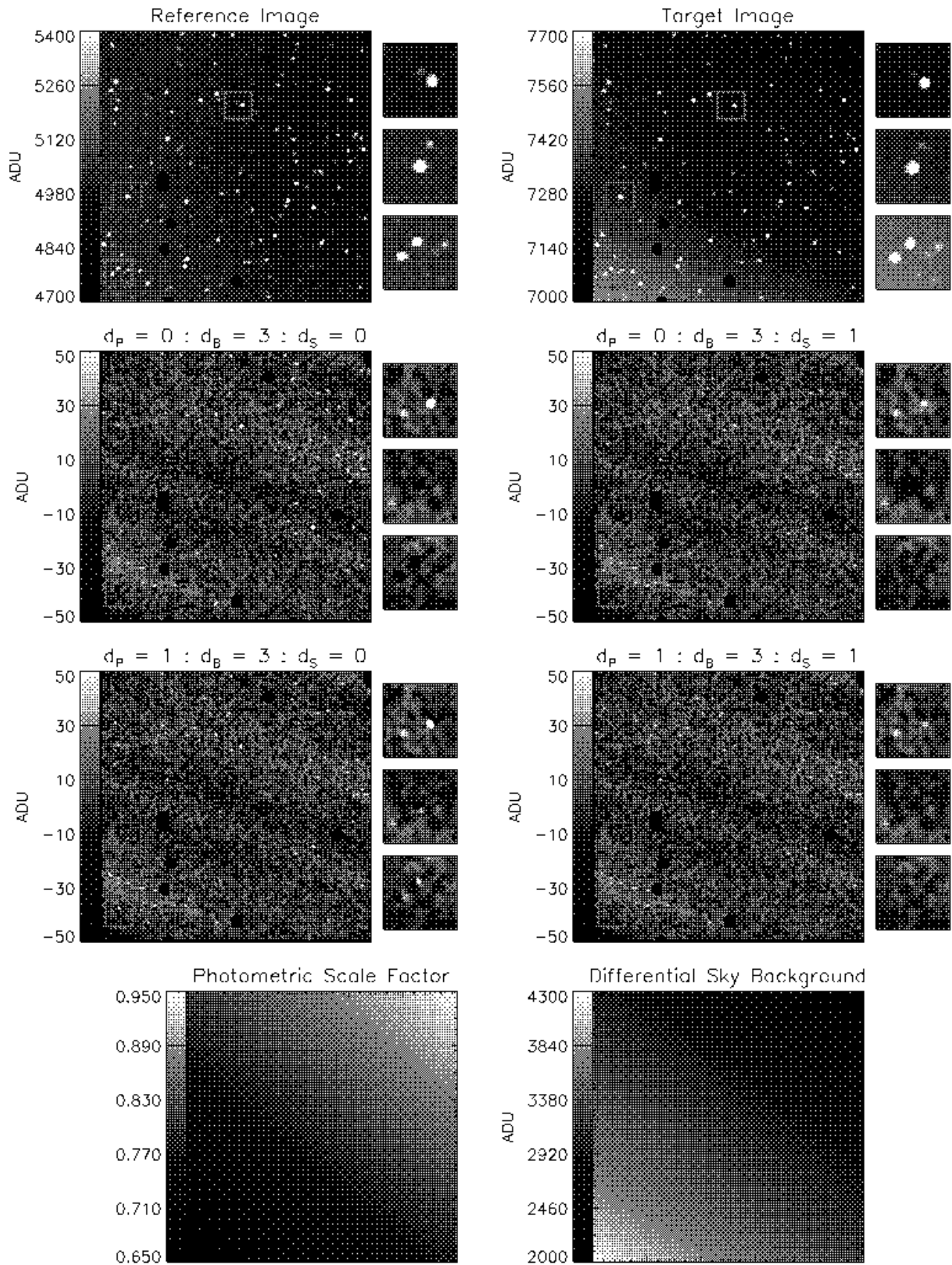,angle=0.0,width=0.97\linewidth}
\caption{Top: A pair of calibrated images where the target image (right) was taken through light clouds. Middle: Difference images
         for various target image models. The red boxes are displayed as magnified image stamps to the right of each
         difference image. Note that in the bottom-left hand corner of the upper image stamp, the positive-negative residuals are caused by a moving object.
         Bottom: The spatial dependence of the fitted photometric scale factor and differential background
         for the target image model with $(d_{\mbox{\scriptsize P}}, d_{\mbox{\scriptsize B}}, d_{\mbox{\scriptsize S}}) = (1,3,1)$.
         See text for more details.
         \label{fig:images}}
\end{figure*}

The target image was chosen specifically because it was taken through light clouds. We cropped the images appropriately in order to
register them to the nearest pixel. We display the reference and target images in the top two
panels of Figure~\ref{fig:images} with the same linear scale and dynamic range of 700~ADU. The black regions are masked pixels that cover
saturated stars. To the right of each image, we show three magnified image stamps corresponding to the red boxes marked in each image.
These image stamps contain some of the brightest stars in the images which are most suitable for inspecting the quality of the difference images
in the following tests.

We proceed to fit the target image using the reference image and a set of delta basis functions representing a square kernel array of size
9$\times$9~pixels. After some experimentation with different values for $d_{\mbox{\scriptsize P}}$, $d_{\mbox{\scriptsize B}}$,
and $d_{\mbox{\scriptsize S}}$, we find that the differential background is only satisfactorily modelled for $d_{\mbox{\scriptsize B}} \ge 3$.
The resulting difference images for each combination of $d_{\mbox{\scriptsize P}}$ and $d_{\mbox{\scriptsize S}}$ taken from the set $\left\{ 0,1 \right\}$
and with $d_{\mbox{\scriptsize B}} = 3$ are displayed in the middle four panels of Figure~\ref{fig:images}, all with the same linear scale.
The complicated residuals in the differential sky background are apparent in all cases.

For $(d_{\mbox{\scriptsize P}}, d_{\mbox{\scriptsize B}}, d_{\mbox{\scriptsize S}}) = (0,3,0)$, the dominant residuals
at the star positions show an under-subtraction of the star fluxes towards the top-right of the difference image, and an over-subtraction
of the star fluxes towards the bottom-left, which is clearly due to the presence of spatial transparency variations that
are not modelled by the spatially invariant photometric scale factor. This is also the case for
$(d_{\mbox{\scriptsize P}}, d_{\mbox{\scriptsize B}}, d_{\mbox{\scriptsize S}}) = (0,3,1)$, but since the kernel model is allowed
to vary in shape across the image area, the zero-sum delta basis functions try to mitigate the spatial transparency variations by
moving flux from the reference image background to the star PSF for those stars whose fluxes are under-subtracted, and by moving
flux from the star PSF to the reference image background for those stars whose fluxes are over-subtracted, resulting in smaller residuals
at the star positions but with the residuals spread out over a larger area. This is most visible in the image stamps on the right which
still show under- and over-subtraction of the star fluxes, but spread out over more pixels. Setting
$(d_{\mbox{\scriptsize P}}, d_{\mbox{\scriptsize B}}, d_{\mbox{\scriptsize S}}) = (1,3,0)$ successfully removes the under- and over-subtraction
of the star fluxes from the difference images, but instead leaves
positive-negative residuals at each star position whose orientation is a function of position, which is a consequence of not modelling
spatial variations in the kernel shape. 

Adopting $(d_{\mbox{\scriptsize P}}, d_{\mbox{\scriptsize B}}, d_{\mbox{\scriptsize S}}) = (1,3,1)$
produces difference images where only the brightest stars can be seen to be mildly under- or over-subtracted, which is a much better result
than what current DIA algorithms are capable of producing (i.e. the $(d_{\mbox{\scriptsize P}}, d_{\mbox{\scriptsize B}}, d_{\mbox{\scriptsize S}}) = (0,3,1)$ case).
It is quite possible that further improvements in the difference image residuals may be obtained by adopting even higher polynomial degrees
for $d_{\mbox{\scriptsize P}}$, $d_{\mbox{\scriptsize B}}$, and $d_{\mbox{\scriptsize S}}$, but a full optimisation of the production
of the difference image in our example is outside of the scope of this paper.

In the bottom two panels of Figure~\ref{fig:images},   
we reproduce the fitted photometric scale factor and differential background as a function of position over the image area
which show that the atmospheric transparency diminishes and the sky background brightens for the parts of the target image
that are more affected by clouds. This result is to be expected since clouds attenuate the incoming light from outside the Earth's atmosphere, but
they also increase the local sky brightness by scattering ambient light (e.g. light pollution, moon light, etc.) back to the ground.

However, to be absolutely sure that this observed anti-correlation is not an artefact of our modelling procedure, we performed the following test.
We cut out ten well-distributed image stamps around bright stars from the reference image, and we also cut out the corresponding
stamps from the target image. For each pair of image stamps, we proceeded to fit the target image stamp using the
reference image stamp and the same kernel configuration that we used to model the full target image, and
we adopted a spatially invariant kernel and differential background (i.e.
$(d_{\mbox{\scriptsize P}}, d_{\mbox{\scriptsize B}}, d_{\mbox{\scriptsize S}}) = (0,0,0)$). We compared the
photometric scale factor and the differential background derived from each fit, which represent robust local
estimates of these quantities, to the predicted values of these quantities at the stamp coordinates from
our model for the full target image, and we found a very good agreement (to within $\sim$2-4 per cent). This confirms that the results from our 
new DIA algorithm are fully consistent with the results that can be obtained using current DIA algorithms.

This real data example has served as a proof-of-concept
where we have demonstrated that we can use our DIA algorithm to successfully model a spatially varying photometric scale factor.
We have also shown how the results of solving for a spatially invariant kernel and differential background for small image sub-regions
(stamps) in different parts of the image can be used to perform consistency checks on the solution for the target
image model from our DIA algorithm.

\section{Implementation Hints And Optimisation Tricks}
\label{sec:impl_and_opt}

Producing difference images is a very computationally intensive task, especially when modelling
a spatially varying kernel as we have described in Section~\ref{sec:general_sol}. However, many
applications of DIA require quick (within seconds or minutes) processing of the target images (e.g. robotic searches for
anomalies in microlensing events towards the Galactic bulge - RoboNet-II - \citealt{tsa2009}, supernovae searches -
Palomar Transient Factory - \citealt{gal2011}, etc.).
Hence the optimisation of the calculations required to produce the difference images is an important
aspect of DIA. In the following subsections, we describe some useful optmisation tricks that may be
used to obtain some substantial improvements in speed, and that have the potential to rival brute force
DIA implementations on graphical processing units (GPUs; \citealt{flu2011}).

\subsection{Memory Considerations}
\label{sec:memory}

Firstly we consider what information needs to be stored in computer memory to enable the
efficient calculation of the least-squares matrix $\mathbf{H}$ and vector $\boldsymbol{\beta}$,
and the target image model $M_{ij}$. We limit ourselves to considering arrays that are of the
size of the target image, since the other information that needs to be stored in computer memory
(e.g. the discrete kernel basis functions $\kappa_{qrs}$) takes up negligible space in comparison.

For efficiency, we need to pre-calculate and store in computer memory those images that will be
used more than once in the calculation of the difference image. In the general case, these images
are the target image $I_{ij}$, the reference image $R_{ij}$, the inverse-variance image $1 / \sigma_{ij}^{2}$,
the $N_{\kappa}$ basis images $[R \otimes \kappa_{q}]_{ij}$, and the $N_{\mbox{\scriptsize poly}}$
polynomial images of the spatial coordinates $\eta^{m}_{i} \, \xi^{n}_{j}$ for $m + n \ge 1$.
If $d_{\mbox{\scriptsize max}}$ is the maximum degree of the polynomial spatial variation of the kernel basis function
coefficients and the differential background, then the maximum degree of the polynomial images of the spatial coordinates
in the least-squares matrix $\mathbf{H}$ is $2 d_{\mbox{\scriptsize max}}$ (see Equation~\ref{eqn:least_squares_matrix}), which implies that:
\begin{equation}
N_{\mbox{\scriptsize poly}} = \left[ (2 d_{\mbox{\scriptsize max}} + 1) (2 d_{\mbox{\scriptsize max}} + 2) / 2 \right] - 1 = d_{\mbox{\scriptsize max}} (2 d_{\mbox{\scriptsize max}} + 3)
\label{eqn:npoly}
\end{equation}

For the Gaussian basis functions, with the typical choice of 53 such functions (see Section~\ref{sec:gauss_basis}), it
is perfectly feasible to store all of the corresponding basis images in computer memory (e.g. 53 floating point
2000$\times$2000 pixel images take up $\sim$831~Mb of memory using {\tt IDL}). Furthermore, the spatial variation of the 
kernel basis function coefficients is not usually modelled with a higher degree polynomial than a cubic polynomial,
and a cubic polynomial variation requires $N_{\mbox{\scriptsize poly}} = 27$. Again, it is possible to store
all of the required $1 + 1 + 1 + 53 + 27 = 83$ images in computer memory (e.g. 83 floating point 2000$\times$2000 pixel images take up
$\sim$1280~Mb of memory using {\tt IDL}).

For the delta basis functions, a typical circular kernel of radius 10~pix generates 349 basis images, which is a more problematic
number of images to store in the computer memory (especially for a 32-bit machine). However, the basis images for the delta basis
functions may be generated without performing a computationally-costly convolution by simply subtracting the reference image from
a shifted version of itself (see Equation~\ref{eqn:delta_basis_images}). Hence, if one is prepared to recalculate each basis image
as needed, and assuming that up to 27 polynomial images are required, then only $1 + 1 + 1 + 1 + 27 = 31$ images need to be stored
in computer memory. Similarly, using the same approach for mixed-resolution delta basis functions with $N_{\mbox{\scriptsize res}}$
resolutions only requires the storage of $N_{\mbox{\scriptsize res}}$ versions of the reference image, each one produced by
convolving the original reference image with a box-car of the shape of the relevant extended delta basis function.

\subsection{Calculating The Least-Squares Matrix}
\label{sec:lsm}

By far, most of the arithmetic operations required to fit the target image model and produce a difference image are performed in
the construction of the least-squares matrix $\mathbf{H}$ and vector $\boldsymbol{\beta}$. In fact,
assuming that the inverse-variance, basis, and polynomial images are pre-calculated, and that $D$ is the polynomial degree of
spatial variation of each kernel basis function and the differential background, then there are 
$N_{\mbox{\scriptsize par}} = (N_{\kappa} + 1) (D + 1) (D + 2) / 2$ coefficients to be determined, and
brute force computation of $\mathbf{H}$ requires the calculation of $N_{\mbox{\scriptsize par}}^{2}$
entries, where the vast majority of these entries require $3 N_{\mbox{\scriptsize pix}}$ multiplications and
$N_{\mbox{\scriptsize pix}} - 1$ additions (note that $N_{\mbox{\scriptsize pix}}$ is the number of target image pixels
that are being modelled). Furthermore, $\boldsymbol{\beta}$ requires the computation of another $N_{\mbox{\scriptsize par}}$ entries, 
where again the vast majority of these entries require $3 N_{\mbox{\scriptsize pix}}$ multiplications and $N_{\mbox{\scriptsize pix}} - 1$
additions. Hence, the number of arithmetic operations $N_{\mbox{\scriptsize op}}$ for the brute force computation of $\mathbf{H}$
and $\boldsymbol{\beta}$, normalised by $N_{\mbox{\scriptsize pix}}$, is given by:
\begin{equation}
N_{\mbox{\scriptsize op}} \approx 4 N_{\mbox{\scriptsize par}} (N_{\mbox{\scriptsize par}} + 1)
\label{eqn:nop_brute}
\end{equation}

We have already mentioned in Section~\ref{sec:input} that limiting the target image pixels to be used in calculating $\mathbf{H}$
and $\boldsymbol{\beta}$ to a set of suitable image sub-regions minimises the number of required arithmetic operations for minimal
loss of precision on the coefficients in the target image model. This clearly follows from the discussion in the previous paragraph.

We have also noted in Section~\ref{sec:fit_target_im_model} that $\mathbf{H}$ is symmetric. This means that in reality only
$N_{\mbox{\scriptsize par}} (N_{\mbox{\scriptsize par}} + 1) / 2$ entries in $\mathbf{H}$ need to be calculated, and that the
number of arithmetic operations reduces to:
\begin{equation}
N_{\mbox{\scriptsize op}} \approx 2 N_{\mbox{\scriptsize par}} (N_{\mbox{\scriptsize par}} + 3)
\label{eqn:nop_sym}
\end{equation}

Now we consider the order in which we may efficiently calculate the entries of $\mathbf{H}$ and $\boldsymbol{\beta}$, and
since $\mathbf{H}$ has the much larger number of entries, our choice is driven by the structure of $\mathbf{H}$.
Note that in the following, we treat the differential background as having a corresponding basis image set to unity
at all pixels (see Section~\ref{sec:target_im_model}). Inspection of
Equation~\ref{eqn:least_squares_matrix} for $\mathbf{H}$ reveals that one has the choice of either:
\begin{enumerate}
\item For each of the $(N_{\mbox{\scriptsize poly}} + 1)$ polynomial images, cycle through the $(N_{\kappa} + 1)^2$ pairs of basis
      images to calculate the corresponding $(N_{\kappa} + 1)^2$ and $N_{\kappa} + 1$ entries in $\mathbf{H}$ and $\boldsymbol{\beta}$, respectively.
\item For each of the $(N_{\kappa} + 1)^2$ pairs of basis images, cycle through the $(N_{\mbox{\scriptsize poly}} + 1)$ polynomial images to calculate the corresponding
      $\left[ (D + 1)(D + 2)/2 \right]^{2}$ and $(D + 1)(D + 2)/2$ entries in $\mathbf{H}$ and $\boldsymbol{\beta}$, respectively.
\end{enumerate}
We note that to calculate each entry in $\mathbf{H}$, a pair of basis images needs to be multiplied together before performing the 
required summation, whereas the polynomial images are already pre-calculated from the coordinate images in computer memory, and therefore
option~(ii) is the most efficient because it minimises the number of the image multiplications that are required.
Furthermore, in the case of the delta basis functions, the basis images are calculated as needed, and therefore
option~(ii) also minimises the number of times that each basis image must be calculated.

\begin{figure}
\centering
\epsfig{file=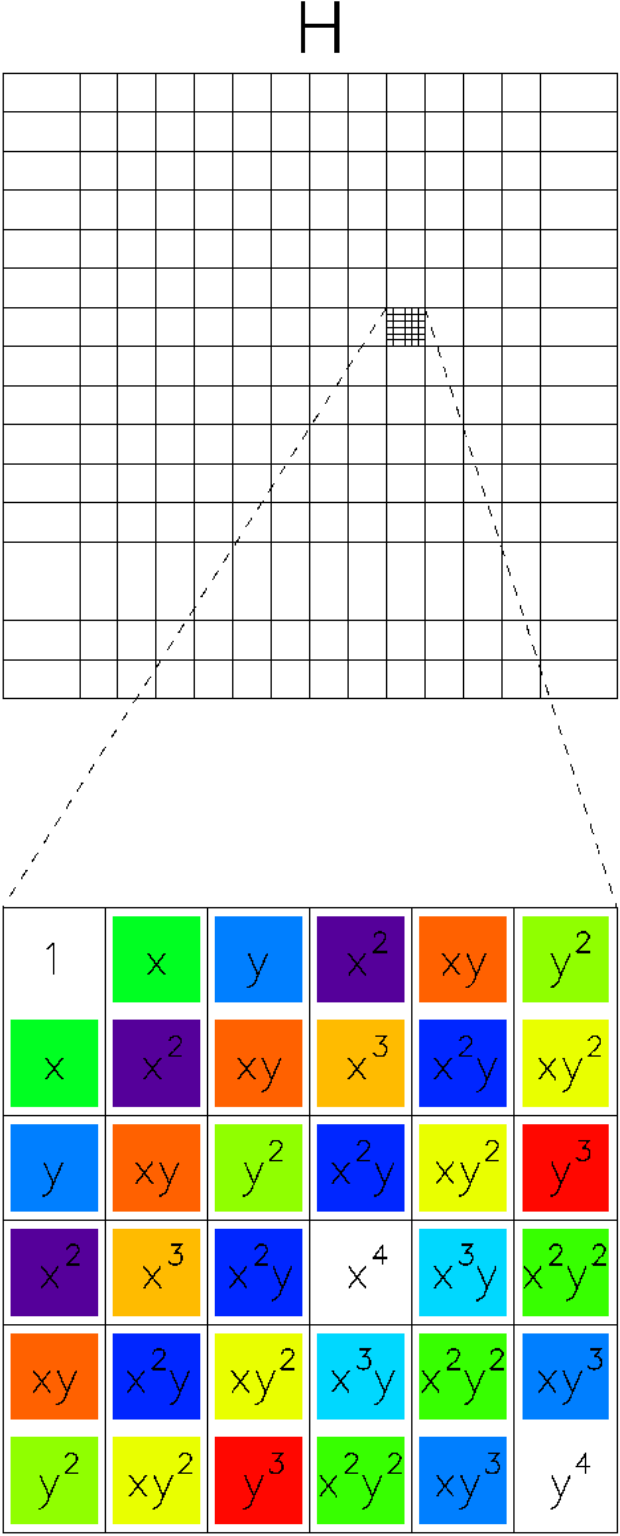,angle=0.0,width=\linewidth}
\caption{Top: The full least-squares matrix $\mathbf{H}$ divided up into $(N_{\kappa} + 1)^2$ square sub-matrices, where
         each sub-matrix has $\left[ (D + 1)(D + 2) / 2 \right]^{2}$ entries. For this example we have adopted an artificially
         small value of $N_{\kappa} = 15$ for clarity, and $D = 2$. Bottom: A magnified view of a single square sub-matrix.
         Each sub-matrix in $\mathbf{H}$ has the same structure. Entries in the sub-matrix that employ the same polynomial
         image in their calculation have the same background colour (except for the single entries corresponding to the
         polynomial images 1, $\eta^{4}_{i}$, and $\xi^{4}_{j}$). The polynomial term marked in each sub-matrix entry indicates
         the degree in the spatial coordinates $(x,y)$ of the polynomial image corresponding to that entry.
         \label{fig:matrix_layout}}
\end{figure}

Having justified the choice of option~(ii) for the order in which we should calculate the entries of $\mathbf{H}$ and $\boldsymbol{\beta}$,
we adopt a corresponding parameter ordering in the parameter vector $\boldsymbol{\alpha}$ that leads to the structure for $\mathbf{H}$
that we illustrate in Figure~\ref{fig:matrix_layout} for $N_{\kappa} = 15$ (artificially small for clarity) and $D = 2$.
The matrix $\mathbf{H}$ is made up of $(N_{\kappa} + 1)^2$ square sub-matrices (top panel of Figure~\ref{fig:matrix_layout}),
where each sub-matrix corresponds to the product of a single basis image pair $[R \otimes \kappa_{q}]_{ij} \, [R \otimes \kappa_{q^{\,\prime}}]_{ij}$.
Furthermore, each square sub-matrix has $\left[ (D + 1)(D + 2) / 2 \right]^{2}$ entries, where each entry corresponds to a different polynomial image.
However, within a single sub-matrix, there are only $N_{\mbox{\scriptsize poly}} + 1 = (D + 1) (2D + 1)$ independent entries
(Equation~\ref{eqn:npoly}; bottom panel of Figure~\ref{fig:matrix_layout}). In our specific example for $D = 2$, there are 15 independent entries out of 36 entries in each
sub-matrix (i.e. less than half of the entries need to be calculated).

The discovery of this property of $\mathbf{H}$ is exceptionally important because it greatly decreases the number of required
calculations. Neither M08 nor \citet{qui2010} mention this optimisation, and A00 claim that the full modelling of the spatial
variation of the kernel ``quickly becomes intractable'', and that ``order 3 requires roughly 100 times more calculations than
a constant kernel solution''. We find that capitalising on the pattern in the sub-matrices of $\mathbf{H}$ for a spatial
variation of the kernel of degree 3, one would only require $N_{\mbox{\scriptsize poly}} + 1 = 28$ times more calculations than
for a constant kernel solution, which is a very significant improvement in the potential performance of the algorithm.

We are now in a position to develop an optimised algorithm for computing $\mathbf{H}$ and $\boldsymbol{\beta}$. We propose the
following procedure:
\begin{enumerate}
\item For each row of square sub-matrices in $\mathbf{H}$, carry out steps (ii)-(vii), and then finish.
\item Calculate $[R \otimes \kappa_{q}]_{ij} \, / \, \sigma_{ij}^{2}$ and $I_{ij} \, [R \otimes \kappa_{q}]_{ij} \, / \, \sigma_{ij}^{2}$
      for the current row, which requires $2 N_{\mbox{\scriptsize pix}}$ multiplications.
\item For each sub-matrix in the current row that lies on the diagonal or in the upper half of $\mathbf{H}$, carry out steps (iv)-(v), and then move
      on to step (vi).
\item Calculate $[R \otimes \kappa_{q}]_{ij} \, [R \otimes \kappa_{q^{\,\prime}}]_{ij} \, / \, \sigma_{ij}^{2}$ for the current sub-matrix,
      which requires $N_{\mbox{\scriptsize pix}}$ multiplications.
\item For each pre-calculated polynomial image, calculate the expression
      $\sum_{ij} \, \eta^{m + m^{\,\prime}}_{i} \, \xi^{n + n^{\,\prime}}_{j} \, [R \otimes \kappa_{q}]_{ij} \, [R \otimes \kappa_{q^{\,\prime}}]_{ij} \, / \, \sigma^{2}_{ij}$,
      which requires $N_{\mbox{\scriptsize pix}}$ multiplications (except for $m + m^{\,\prime} + n + n^{\,\prime} = 0$)
      and $N_{\mbox{\scriptsize pix}} - 1$ additions, and fill out the relevant entries of the current sub-matrix.
\item Fill out the entries of the sub-matrices in the current row that lie in the lower half of $\mathbf{H}$ by using the fact
      that $\mathbf{H}$ is symmetric, which takes a negligible number of operations.
\item For each relevant pre-calculated polynomial image, calculate the expression
      $\sum_{ij} \, \eta^{m}_{i} \, \xi^{n}_{j} \, I_{ij} \, [R \otimes \kappa_{q}]_{ij} \, / \, \sigma^{2}_{ij}$, which requires $N_{\mbox{\scriptsize pix}}$ 
      multiplications (except for $m + n = 0$) and $N_{\mbox{\scriptsize pix}} - 1$ additions, and fill out the corresponding entries in $\boldsymbol{\beta}$.
\end{enumerate}

\begin{figure}
\centering
\epsfig{file=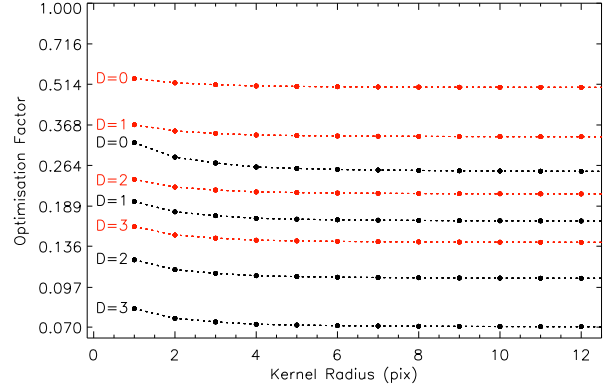,angle=0.0,width=\linewidth}
\caption{Plot of the ratio of the number of arithmetic operations required to calculate $\mathbf{H}$ and $\boldsymbol{\beta}$
         for our optimised algorithm compared to the same quantity for the brute force computation (black),
         and for the brute force computation that capitalises on the symmetry in $\mathbf{H}$ (red), when we adopt
         a set of delta basis functions representing a circular kernel. These ratios are plotted as a function of the
         kernel radius (pix) and for $D =$0, 1, 2, and 3.
         \label{fig:algo_perf}}
\end{figure}

We now attempt to estimate the number of arithmetic operations that are required to calculate $\mathbf{H}$ and
$\boldsymbol{\beta}$ using our optimised algorithm. Observe that step (ii) is repeated $N_{\kappa} + 1$ times,
steps (iv) and (v) are each repeated $(N_{\kappa} + 1) (N_{\kappa} + 2) / 2$ times of which step (v) requires
$\sim (2 N_{\mbox{\scriptsize pix}}) N_{\mbox{\scriptsize poly}} + N_{\mbox{\scriptsize pix}}$ arithmetic operations, and step (vii)
is repeated $N_{\kappa} + 1$ times and requires $\sim (2 N_{\mbox{\scriptsize pix}}) \left[ (D + 1) (D + 2) / 2 \right] - N_{\mbox{\scriptsize pix}}$ arithmetic operations.
Using $N_{\mbox{\scriptsize poly}} = D \, (2D + 3)$, then we derive the number of arithmetic operations
in our optimised algorithm, normalised by $N_{\mbox{\scriptsize pix}}$, to be:
\begin{equation}
N_{\mbox{\scriptsize op}} \approx (N_{\kappa} + 1) \left[ N_{\kappa} (D + 1)(2D + 1) + 5D^{2} + 9D + 5 \right]
\label{eqn:nop_opt}
\end{equation}

In Figure~\ref{fig:algo_perf}, for $D =$0, 1, 2, and 3, we plot in black the ratio of the expression in Equation~\ref{eqn:nop_opt}
to the expression in Equation~\ref{eqn:nop_brute}
as a function of the kernel radius (pix) for a set of delta basis functions representing a circular kernel.
We see that for typical kernel radii of $\sim$8-12~pix, we expect that our optimised algorithm will reach an efficiency in
the number of arithmetic operations of $\sim$0.251, 0.167, 0.104, and 0.070 compared to the brute force computation for $D =$0, 1, 2, and 3, respectively. Also, in 
Figure~\ref{fig:algo_perf}, for $D =$0, 1, 2, and 3, we plot in
red the ratio of the expression in Equation~\ref{eqn:nop_opt} to the expression in Equation~\ref{eqn:nop_sym} as a function of the kernel radius (pix)
for the same set of delta basis functions.
We further conclude that our optimised algorithm will reach an efficiency in the number of arithmetic operations of
$\sim$0.501, 0.334, 0.209, and 0.140 compared to the brute force computation
that capitalises on the symmetry in $\mathbf{H}$ for $D =$0, 1, 2, and 3, respectively.

\section{Conclusions}
\label{sec:conclusions}

The general framework presented in this paper treats the problem of matching the PSF, photometric scaling, and sky background
between two images, where each of these components varies as a polynomial of the spatial coordinates. Where this paper improves over previous
works on DIA are as follows:
\begin{itemize}
\item{We demonstrate how to model a spatially varying photometric scale factor within our framework,
      which is a new concept that will be important for DIA applied to wide-field
      imaging data that may suffer transparency and airmass variations across the field-of-view.}
\item{We show how to decouple the spatial variation of each kernel basis function, the photometric scale factor, and the differential
      background from each other, which allows more control over the level of spatial variation of each component in the target image model.}
\item{In Section~\ref{sec:general_sol} we develop what we hope is a clear notation and logical order for the DIA equations and methodology aimed at
      aiding others in creating DIA software implementations.}
\item{We prove the equivalence of adopting delta basis functions for the kernel model and solving directly for the kernel pixel values (B08).}
\item{We introduce the mixed-resolution delta basis functions with the aim of reducing the size of the least-squares
      problem to be solved when using delta basis functions, and we elucidate their properties and implications for DIA.}
\item{We present some important optimisations in the calculation of the least-squares matrix which lead to a reduction in the number
      of arithmetic operations that need to be performed for typical kernel radii of $\sim$8-12~pix to $\sim$16.7\%, $\sim$10.4\%, and $\sim$7.0\%
      compared to the brute force computation for linear, quadratic, and cubic spatial variations, respectively, of the target
      image model.}
\end{itemize}

\section*{Acknowledgements}

Dedicated to my three stars $\boldsymbol{\bigstar} \boldsymbol{\star} \boldsymbol{\star}$.

The research leading to these results has received funding from the   
European Union Seventh Framework Programme (FP7/2007-2013) under grant
agreement numbers 229517 and 268421.
We thank the Qatar Foundation for support via QNRF grant NPRP-09-476-1-78.

\section*{Appendix A}
\label{app:append_A}

Here we show that the convolution of the reference image $R_{ij}$ with a continuous
kernel basis function $\kappa_{q}(u,v)$ may be calculated as a discrete convolution.

Firstly, consider the definition of continuous convolution applied to the convolution
of the reference image:
\begin{equation}
[R \otimes \kappa_{q}](x,y) = \int_{-\infty}^{\infty} \int_{-\infty}^{\infty} \! R(x+u,y+v) \, \kappa_{q}(u,v) \,\, \text{d}u \,\, \text{d}v
\label{eqn:append_A1}
\end{equation}
where $R(x,y)$ is a continuous representation of the reference image.

Over the area of one pixel with coordinates $(x_{i},y_{j})$, the value of the reference image is a constant, i.e.
$R(x,y) = R_{ij}$ for \\
$x_{i} - 1/2 \le x < x_{i} + 1/2$ and $y_{j} - 1/2 \le y < y_{j} + 1/2$, and therefore:
\begin{equation}
[R \otimes \kappa_{q}](x_{i},y_{j}) = \sum_{rs} R_{(i+r)(j+s)}
                      \int_{s - \frac{1}{2}}^{s + \frac{1}{2}} \int_{r - \frac{1}{2}}^{r + \frac{1}{2}} \! \kappa_{q}(u,v) \,\, \text{d}u \,\, \text{d}v
\label{eqn:append_A2}
\end{equation}
where $r$ and $s$ are integer indices varying over the domain where the kernel basis function achieves non-zero values.

Adopting the notation $[R \otimes \kappa_{q}]_{ij}$ for the image $[R \otimes \kappa_{q}](x_{i},y_{j})$, then we may write:
\begin{equation}
[R \otimes \kappa_{q}]_{ij} = \sum_{rs} R_{(i+r)(j+s)} \, \kappa_{qrs}
\label{eqn:append_A3}
\end{equation}
\begin{equation}
\kappa_{qrs} = \int_{s - \frac{1}{2}}^{s + \frac{1}{2}} \int_{r - \frac{1}{2}}^{r + \frac{1}{2}} \! \kappa_{q}(u,v) \,\, \text{d}u \,\, \text{d}v
\label{eqn:append_A4}
\end{equation}
where $r$ and $s$ now represent the pixel indices corresponding to the column $r$ and row $s$ of the discrete kernel
basis function $\kappa_{qrs}$.

Hence, the image $[R \otimes \kappa_{q}](x,y) = [R \otimes \kappa_{q}]_{ij}$, which we refer
to as a {\it basis image}, may be calculated via the
discrete convolution defined in Equation~\ref{eqn:append_A3}.

\section*{Appendix B}
\label{app:append_B}

We wish to briefly investigate the consequences of approximating $M_{ij}$ with $I_{ij}$ in the noise model
in Equation~\ref{eqn:noise_model} as opposed to iterating the solution and using the current image model from
Equation~\ref{eqn:model_lincomb_discrete} to update the noise model at each iteration. For this purpose we
use the software developed in B08 for the case of a kernel and differential background
that are both spatially invariant.

We create a 205$\times$205~pixel noiseless reference image $R_{ij}$ by setting a constant sky level of 1000~ADU and adding in
100 objects, each of flux 10$^{5}$~ADU and with a two-dimensional Gaussian profile of FWHM 4~pix,
at random spatial coordinates drawn from a uniform distribution across the image area.
We also create a 201$\times$201~pixel noiseless target image $S_{ij}$ by convolving the $R_{ij}$ with a discrete 5$\times$5~pixel
kernel calculated via numerical integration of Equation~\ref{eqn:discrete_ker_basis_func} for a two-dimensional Gaussian
of FWHM 2~pix centred at the kernel centre and normalised to a sum of unity.

We then perform the following experiment, adopting reasonable values for the readout noise and gain of $\sigma_{0} = 5$~ADU
and $G = 1$~e$^{-}$/ADU, respectively:
\begin{enumerate}
\item We generate a 201$\times$201~pixel image $\Sigma_{ij}$ of values drawn from a normal distribution with zero mean
      and unit $\sigma$, and we construct a noisy target image $I_{ij}$ via:
      \begin{equation}
      I_{ij} = S_{ij} + \Sigma_{ij} \sqrt{ \sigma_{0}^{2} + S_{ij} }
      \end{equation}
      where the coefficient of $\Sigma_{ij}$ is derived from Equation~\ref{eqn:noise_model} for $G = 1$~e$^{-}$/ADU and $F_{ij} = 1$.
\item We solve for a kernel and differential background that are both spatially invariant to match the reference image $R_{ij}$ to the
      target image $I_{ij}$. For the kernel model, we adopt 25 delta basis functions covering a 5$\times$5~pixel array to
      match the actual domain of the discrete pixel kernel used to generate $S_{ij}$ from $R_{ij}$. For the target image noise model
      $\sigma_{ij}$ we use Equation~\ref{eqn:noise_model} with $M_{ij}$ approximated by $I_{ij}$.
\item We record the photometric scale factor $P_{\,1}$ and differential background $B_{1}$ of the solution obtained in step~(ii).
\item We iterate the solution for the spatially invariant kernel and differential background three times (sufficient for
      convergence), each time using the current image model $M_{ij}$ calculated via Equation~\ref{eqn:model_lincomb_discrete} to
      set the target image noise model $\sigma_{ij}$ via Equation~\ref{eqn:noise_model}.
\item Again we record the photometric scale factor $P_{\,2}$ and differential background $B_{2}$ of the solution obtained during the final
      iteration in step~(iv).
\end{enumerate}

We repeat the above experiment 10$^{5}$ times and calculate the mean and standard deviation of each of the 
quantities $P_{\,1}$, $B_{1}$, $P_{\,2}$, and $B_{2}$. We find that $\left< P_{\,1} \right> - 1 = 5.38 \times 10^{-6} \pm 1.68 \times 10^{-6}$ and
$\left< B_{1} \right> = -1.0085 \pm 0.0020$~ADU, where the
uncertainty in the mean is estimated from the standard deviation divided by $\sqrt{10^{5}}$. The correct
solution in our experiment should have a photometric scale factor of unity and a differential background of zero. Clearly,
solving the DIA problem using the data to estimate the uncertainties on the pixel values in the target image introduces
a bias of $\sim$1~ADU in the differential background (and a very slight bias in the photometric scale factor). Hence
one cannot assume that the background in the difference images produced using this method is zero, and aperture photometry
on such difference images should include the computation and subtraction of a local background, and PSF photometry should
include the local background as a parameter in the fit. The bias in the differential background solution, which corresponds to an underestimated 
sky background in the target image model, is easily explained by the fact that the background pixels in the target image
that randomly have smaller values than the true sky background are given more weight (or smaller uncertainties) in the fit than
those background pixels that randomly have larger values than the true sky background.

For the case where we iteratively solve the DIA problem using the current image model to determine the uncertainties on
the target image pixel values at each iteration, we find that $\left< P_{\,2} \right> - 1 = 1.98 \times 10^{-6} \pm 1.68 \times 10^{-6}$
and $\left< B_{2} \right> = -0.0031 \pm 0.0020$~ADU. Therefore, at the precision of our experiment (which is well beyond the
photometric precision typically obtained for real data), we conclude that there is no bias in the
derived photometric scale factor or differential background for this method, which validates the iterative method
presented in Section~\ref{sec:uncert_and_iter}.

Finally we mention that even though we only report one particular experiment in this Appendix, we actually performed a range
of experiments on artificial noisy target images generated with different set-ups (e.g. different convolution kernels)
and we found similar results in all cases.

\label{lastpage}

\end{document}